\documentclass[pra,floatfix,twocolumn,showpacs]{revtex4}
\usepackage{dcolumn}
\usepackage{bm}
\usepackage{epsfig}
\usepackage{color}

\begin{document}
\title{
Solitons and Josephson-type oscillations in Bose-Einstein condensates with spin-orbit coupling and 
time-varying Raman frequency}
\author{F. Kh. Abdullaev$^{1,2}$, M. Brtka$^{2}$, A. Gammal$^{3}$, Lauro Tomio$^{4,5}$}
\affiliation{$^{1}$Physical-Technical Institute, Uzbekistan Academy of Sciences, Tashkent, Bodomzor yuli, 2-b, Uzbekistan}
\affiliation{$^{2}$CCNH e CMCC, Universidade Federal do ABC, 09210-170, Santo Andr\'e, Brazil}
\affiliation{$^{3}$Instituto de F\'isica, Universidade de S\~{a}o Paulo, 05508-090, S\~{a}o Paulo, Brazil}
\affiliation{$^{4}$Instituto Tecnol\'ogico de Aeron\'autica, CTA, 12228-900, S\~ao Jos\'e dos Campos, Brazil}
\affiliation{$^{5}$Instituto de F\'\i sica Te\'orica, Universidade Estadual Paulista, 01140-070, S\~{a}o Paulo, Brazil}
\date{\today}
\begin{abstract}
The existence and dynamics of solitons in quasi-one-dimensional 
Bose-Einstein condensates (BEC) with spin-orbit coupling (SOC) and attractive two-body 
interactions  are described for two coupled atomic pseudo-spin components with slowly and 
rapidly varying time-dependent Raman frequency.  
By varying the Raman frequency linearly in time, it was shown that ordinary nonlinear Schr\"odinger-type 
bright solitons can be converted to striped bright solitons and vice versa.  
The internal Josephson oscillations between atom-number of the coupled soliton components, and the corresponding
center-of-mass motion,  are studied for different parameter configurations. In this case, a mechanism to control the soliton 
parameters is proposed by considering parametric resonances, which can emerge when using time-varying Raman frequencies.
Full numerical simulations confirm variational analysis predictions when applied to the region where regular solitons are 
expected. In the limit of high frequencies, the system is described by a time-averaged Gross-Pitaevskii formalism 
with renormalized nonlinear and SOC parameters and modified phase-dependent nonlinearities. 
By comparing full-numerical simulations with averaged results, we have also studied the lower limits for the frequency of 
Raman oscillations in order to obtain stable soliton solutions. 
\end{abstract}
\maketitle

\section{Introduction}
A progressively growing interest in the physics of Bose-Einstein condensate (BEC) with spin-orbit coupling (SOC) has been 
observed in recent years~\cite{2015Zhai,2011LJS,LC2009,2011Ho,2017Li}. 
For the coupling, different forms of  Rashba~\cite{BR84} and Dresselhaus~\cite{Dresselhaus55}, as well as mixture of them, have been
realized~\cite{2013Galitskii}.  Important forms for the nonlinear excitations have been verified with structure of
stable solitons, for condensed systems having attractive and repulsive interactions. In this regard, we can mention
that  the existence of solitons in BEC with SOC was investigated in Ref.~\cite{Merkl10}, for the case that we have 
repulsive interactions between atoms; and, in Refs.~\cite{ZPZ13,AFKP,DKM,SM13}, when the interactions are 
attractive. Gap solitons are predicted in Ref.~\cite{KKA-PRL13} for BEC with SOC in a spatially periodic
Zeeman field, corresponding to a linear optical lattice. Experimentally, it is not an easy task to control the SOC 
parameter, having recent suggestions to tune it by applying rapid time variations of the Raman 
frequency~\cite{2013Zhang,Salerno2016}.  The experimental observations reported in Ref.~\cite{2015Spielman} 
are shown that the spin-orbit coupling can be tuned in this way. 
In principle, the periodic variation in time of the condensate parameters can lead to new phenomena as the generation 
of new quantum phases~\cite{Struck}, artificial gauge fields~\cite{2016FE}, compacton matter waves~\cite{AKS}, etc.
Therefore, actually it is quite relevant and of interest to investigate how the periodic time variations of the Raman 
frequency can affect the nonlinear modes of the condensate, such as solitons and vortices. In the limit of high 
frequencies, as shown in Ref.~\cite{2013Zhang}, the averaged Hamiltonian contains the nontrivial renormalization 
of the spin-orbit coupling, as well as the new effective nonlinear phase, which is sensitive to the interaction
strengths~\cite{SKMA}. 

Our main task in the present work is to investigate the dynamics of solitons and  Josephson-type oscillations between 
solitonic components, considering BEC with  tunable spin-orbit coupling, with attractive interactions between the atoms, 
under slow and rapid time modulations of the Raman frequency. 
In this regard, related to Josephson oscillations in BEC, we should mention some previous studies in  Refs.~\cite{albiez2005,levy2007,abbarchi2013,2014zou}.  
In particular, when considering the Raman frequency modulated in time, together with changes in other parameters
of the system, one should expect to observe parametric resonance phenomena to occur in the internal Josephson oscillations,
which have been introduced between the atom-number fraction existent in each of the components of the condensate 
with SOC.
The parametric resonances in this case are introduced by the time-dependence of the Raman frequency 
and its corresponding relation with spin-orbit coupling of the two components of the condensate.  
Such study can be useful for possible experimental  investigations, which can help the control of BEC parameters through 
resonance phenomena observations. 
We should point out that previous studies on parametric resonances in BEC are mainly concerned with 
time variations in trap configurations, optical lattices, as well as nonlinear parameters, looking for direct interference
effects manifested in the densities~\cite{1999ripoll,2000kevrekidis,2002salasnich,2005tozzo,2014cairncross,2016abdullaev}. 
In our present study, by introducing a time modulation in the Raman frequency, the main focus is the oscillatory behavior 
between the internal atom-number population of the two components, during the time evolution of the condensate.

We start our study by first considering the case that we have defined spin-orbit parameter and constant Raman 
frequency, in order to verify the characteristics of the existent soliton solutions, which can be regular or
striped solitons. Next, we introduce an adiabatic time modulation in the Raman frequency (growing and decreasing
linearly), such that we can study how to switch between different kind of solitons. We follow our study by considering 
the Josephson oscillations between the components for both the cases that we have regular and striped soliton solutions.
Parametric resonance effects in the oscillations are verified by considering periodically time variation of the Raman 
frequency at some given SOC parameters.
The case of rapid time modulations of the Raman frequency can be treated by using a time-averaging approach, which is 
implemented over the time-dependent coupled system, implying in a renormalization of the SOC and nonlinear parameters. 
In this way, the interactions are effectively time independent. This case is being discussed in the final part (section IV)
of the present work.

We are considering exact numerical simulations in all the cases, complemented by theoretical analysis, using 
variational approaches, whenever simplified solutions can be performed. 
As shown, the predictions derived by using the variational approach (VA) are verified to be fully consistent with the given 
numerical results in the region where regular bright solitons are obtained. In other cases, where the solutions are 
striped ones, demanding more parameters in the ansatz, the VA is quite helpful to indicate regions of stability, as well
as initial starting profiles for the full numerical computation. 
By using  the multi-scale expansion method for the averaged system, solitonic solutions are also found and confirmed 
by our numerical simulations of the full coupled system with time-dependent Raman frequency.

In the next, the basic formalism of the model is being presented in section II. For reference to other sections, we add two
subsections: The first (A) where we provide some details on the linear energy dispersion, which is defining two regions for the 
kind of soliton solutions that we can obtain (regular or striped). In the subsection B, we add already some results to exemplify 
the two possible solutions and how to transform solitons between the two regions by considering 
adiabatic linear time variation of the Raman frequency. In section III, by considering Raman frequency modulated in time and
Josephson oscillations, we analyze the possibility to obtain resonant responses.  The case with Raman having high frequency 
modulations is presented in section IV. Finally, in section IV, we present our main conclusion.

\section{Model formalism}

In our approach, we consider a spin-orbit coupled BEC with equal Rashba and Dresselhaus contributions for the 
spin-orbit coupling  terms, as in Ref.~\cite{AFKP}, which can be described by a one-dimensional (1D) coupled equation 
for the two pseudo-spin components. For that, let us consider an harmonic trap where the frequency along one direction, 
$\omega_x$, is much smaller than the frequency in the perpendicular direction, $\omega_\perp$. In this case, 
given the units of energy, length and time, respectively, by $\hbar \omega_{\perp}$, $a_{\perp}=\sqrt{\frac{\hbar}{m\omega_{\perp}}}$ 
and $1/\omega_{\perp}$ (where $m$ is the mass of both atomic components), we can write in dimensionless units the corresponding 
SOC formalism for the two pseudo-spin components, $u\equiv u(x,t)$ and $v\equiv  v(x,t)$, of the total wave function 
\begin{equation}\label{eq01}
\psi\equiv \psi(x,t) \equiv \left( \begin{array}{c} u \\ v \end{array} \right).
\end{equation}
The corresponding matrix-formatted non-linear Schr\"odinger type coupled equation can be written as
{\small \begin{eqnarray} 
\mathrm{i}\frac{\partial }{\partial t} 
\left( \begin{array}{c} u \\ v \end{array} \right)
&=& \left[-\frac{1}{2}\frac{\partial^2 }{\partial x^2} -\mathrm{i}k_L
\sigma_z\frac{\partial }{\partial x} + V_{tr}+
\Omega(t)\sigma_x \right]
\left( \begin{array}{c} u \\ v\end{array} \right) \nonumber\\
&- & 
 \left( \begin{array}{cc} |u|^2 + \beta|v|^2&0 \\ 0&
 \beta |u|^2 + \gamma|v|^2 \end{array} \right)\left( \begin{array}{c} u \\ v\end{array} \right),
  \label{eq02}\end{eqnarray}}
where $V_{tr}\equiv V_{tr}(x)\equiv \left({\omega_x}/{\omega_\perp}\right)^2 x^2/{2}$ is the trap potential,
assumed to be zero ($V_{tr}=0$) in the present study. $\sigma_{x,z}$ are the usual Pauli matrices, with 
$k_L$ being the strength of the spin-orbit coupling and $\Omega(t)$ the time-dependent Raman frequency 
(also given in units of the trap frequency $\omega_\perp$).
In the non-linear terms we have the dimensionless parameters $\beta$ and $\gamma$, which are given by 
the ratio between the two-body scattering lengths, $a_{ij}$ $(i,j=1,2)$,  of the two atomic components, such that
$\beta =\left|{a_{12}}/{a_{11}}\right|, \gamma=\left|{a_{22}}/{a_{11}}\right|$.  In the present work, we are going to
consider attractive two-body interactions, such that we have an overall minus signal for the non-linear 
interaction. From the symmetry of the coupled equations (\ref{eq02}) for $\gamma=1$, 
we can extract a simple relation between the two components $u$ and $v$:  
Let us consider that, for a given parameter $k_L$ we are identifying the solution for $u$ by
$u(x,t)\equiv u_{(k_L)}(x,t)$. In this case, a particular solution decoupling the equations can be verified with 
$v(x,t)=\pm u_{(-k_L)}(x,t)$.

\subsection{Linear energy spectrum }
\begin{figure}[tbph]
\centerline{
\includegraphics[width=8cm,height=8cm]{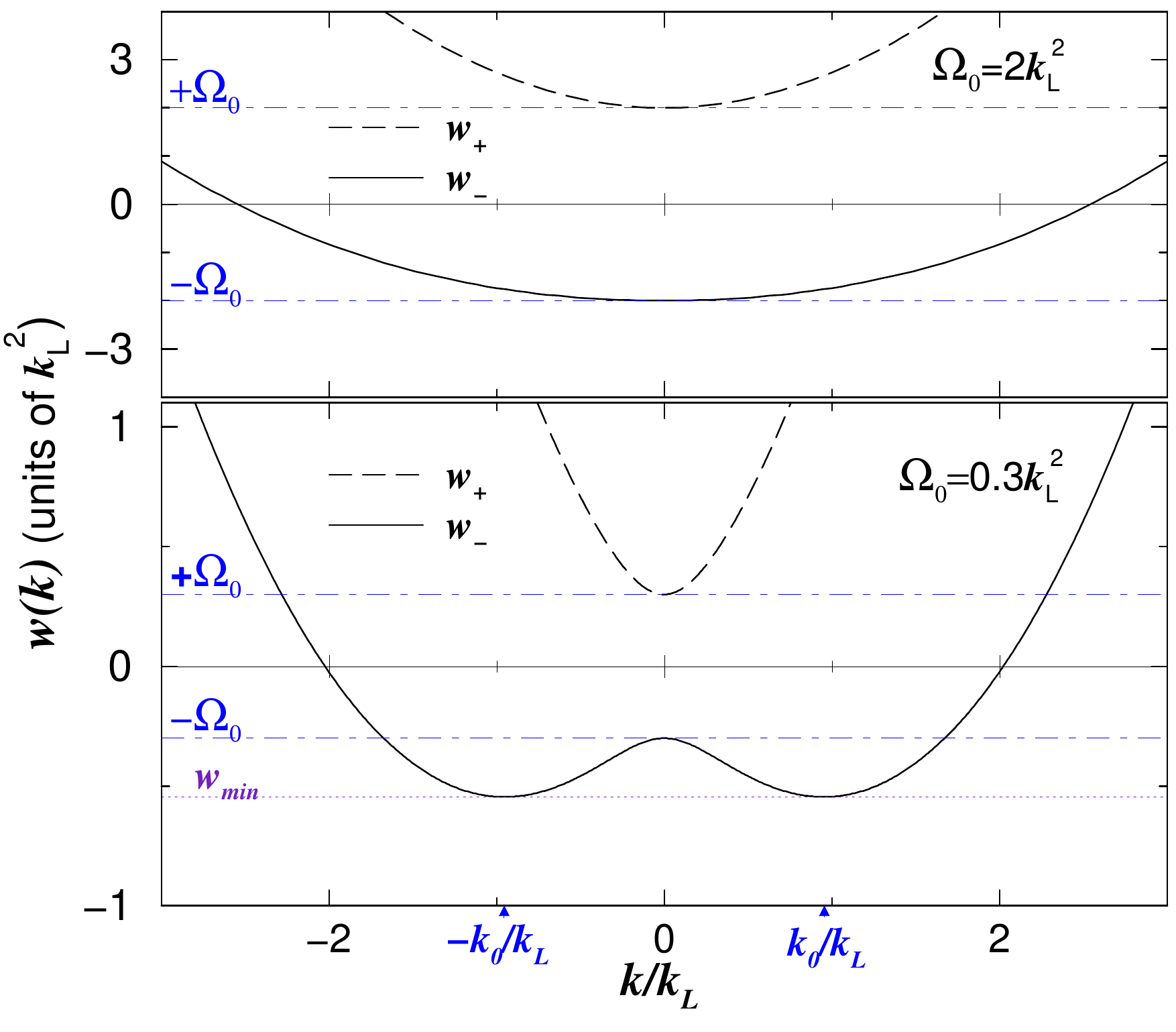}}
\caption{ 
Energy dispersions, $w_{+}(k)$ (dashed curves) and $w_{-}(k)$ (solid curves), given by (\ref{eq03}), 
are shown for two regions of parameters:
Region I, when $k_{L}^2 < \Omega_0$ (upper panel, exemplified with $\Omega_0=2k_L^2$); and 
region II, when $k_{L}^2 > \Omega_0$ (lower panel, exemplified with $\Omega_0=0.3k_L^2$).
In both panels, the dot-dashed and dotted lines indicate the extremes, with the minima of $w_{+}$ being at
$(k,w_+)=(0,\Omega_0)$. The minima of $w_{-}$ are at $(k,w_-)=(0,-\Omega_0)$ in region I;
and at $(\pm k_0,w_{min})$ in region II, where $k_0^2=k_L^2-\Omega_0^2/k_L^2$ and 
$w_{min}=-k_L^2+k_0^2/2$. 
}
\label{fig-01}
\end{figure}

For a constant Raman frequency $\Omega(t)=\Omega_0$, the linear energy spectrum can be derived
by  considering a plane wave-function with momentum $k$,
$\left(u,v\right)=\left(u_0,v_0\right)\exp[{\rm i}(kx-w(k)t)]$, which will give us the following 
dispersion relation:
\begin{equation}\label{eq03}
w_\pm(k)=\frac{1}{2}{k^2} \pm \sqrt{k_{L}^2 k^2 + \Omega_0^2}.
\end{equation}
This relation, also shown in Ref.~\cite{AFKP}, is plotted as a function of $k$ in Fig.~\ref{fig-01}, where  two 
different regions (I and II) can be distinguished, according to the choice of parameters we have
for the spin-orbit coupling $k_L$ and Raman frequency $\Omega_0$.

In region I, which happens when $k_{L}^2 < \Omega_0$, we can only obtain two single minima in the 
dispersion relation; when $w_\pm(k) = w_\pm(0) = \pm \Omega_0$. 
For $k_L^2 > \Omega_0 $, we are in region II, where $w_+(k)$ has just one minimum (at $k=0$, as in the 
case of region I). However, in this case, the dispersion relation for $w_-(k)$ presents a local maximum at $k=0$ 
with two minima at $k=\pm k_0$, where $k_0\equiv k_L\sqrt{1-\Omega_0^2/k_L^4}$; both with 
$w_{min}=k_0^2/2-k_L^2=-\left(k_L^2+\Omega_0^2/k_L^2\right)/2$. 
 
As already verified in Ref.~\cite{AFKP}, bright soliton solutions of the NLSE are obtained in region I. However, 
for $k_L^2>\Omega_0$ (region II), the solutions are striped-type bright solitons. 
In Fig.~\ref{fig-01}, the two regions are represented, exemplified for particular values of $\Omega_0$,  $=2k_L^2$ (region I)
and $=0.3k_L^2$ (region II). The bright solitonic solutions can be found by the multiple scale analysis, which 
will be used in section IV to investigate the soliton dynamics under rapid modulations of parameters.
Here, we should remark that observation of striped phases have been recently reported in Ref.~\cite{2017Li} 
for BEC with SOC.
In the following subsection, we are exemplifying the kind of soliton solutions we obtain in both regions, using constant and 
linearly time-varying Raman frequencies.

\subsection{Regular and striped soliton interchanged by adiabatic time variation of $\Omega(t)$}
\begin{figure}[tbph]
\centerline{
\includegraphics[width=8.5cm]{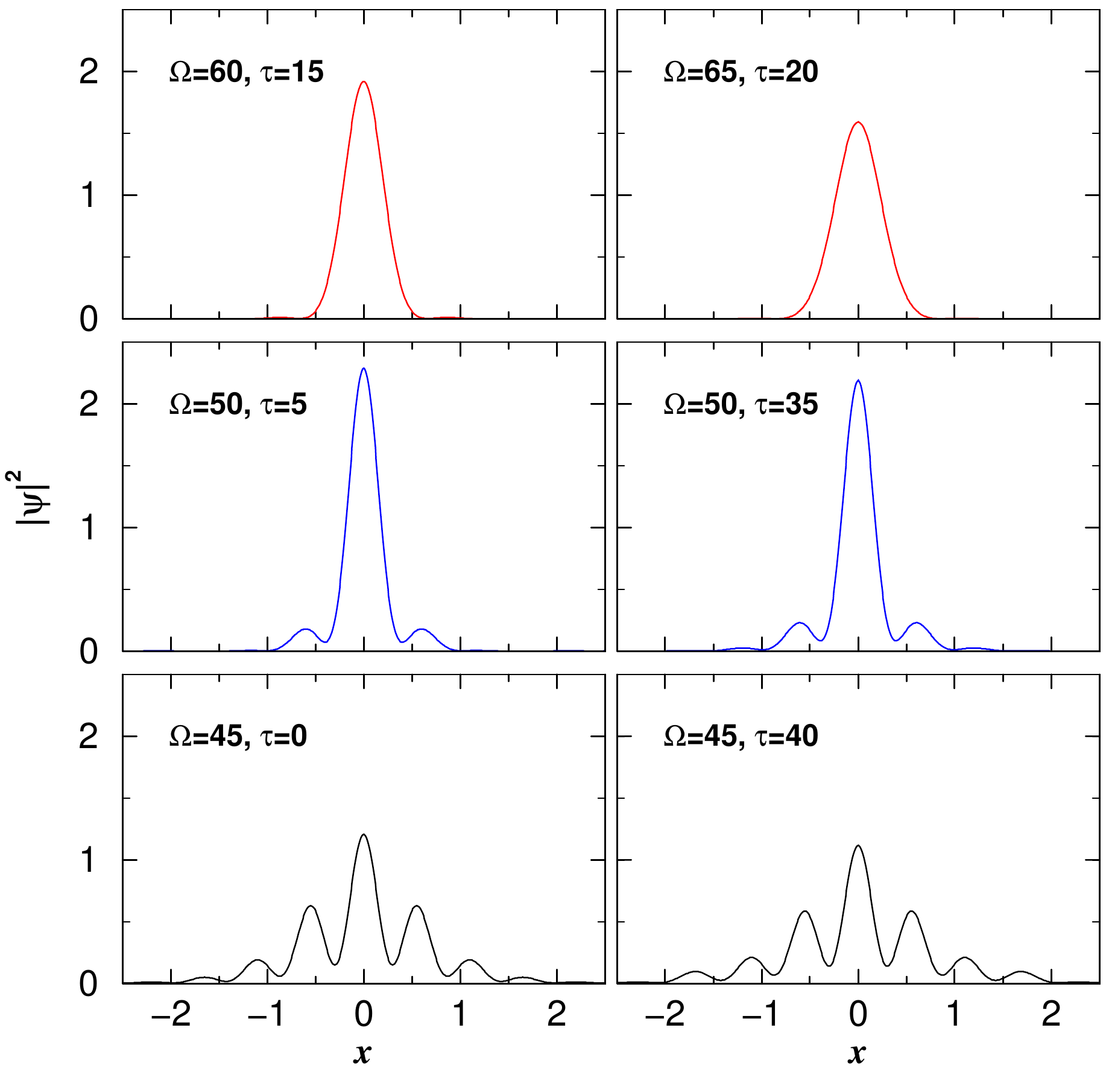}}
\caption{
A soliton profile is shown in the above set of panels, for $\Omega(t)$ varying in time as $\Omega(\tau)=(45+\tau)\Theta(20-\tau)+(85-\tau)\Theta(\tau-20)$ [where $\Theta(x)= 0$ for $x<0$, and 1 for $x>0$]. The parameters are $k_L=8$, $\beta=1$ and $\gamma=1$.
The soliton is generated as a striped one (region II, lower-left panel with $\Omega=45$), being converted to a regular soliton (region I, in the upper-right panel with $\Omega=65$); returning to region II by inverting the time variation. All quantities are dimensionless.
}
\label{fig-02}
\end{figure}

With adiabatic time variation of the Raman frequency, $\Omega(t)$, we can transform solitons from one region
to another region, for a given fixed $k_L$. By considering a regular soliton obtained in region I for $\Omega(t)=\Omega_0$, it can be 
transformed to a striped soliton by decreasing $\Omega(t)$; or, the other way, if started in region II. 
For that, let us consider a variation of the form
\begin{equation}
\Omega(t)= \Omega_0(1 \pm \Delta t).
\label{eq04}\end{equation}
Recently such variation has been used in  numerical simulations for dark soliton generations in BEC with SOC~\cite{josa17}.
The transition of a soliton solution obtained in region II (striped soliton) to the region I (regular soliton), and back from 
region I to region II, is illustrated in Fig.~\ref{fig-02}, by considering several panels,  obtained for different values of a 
the time-dependent Raman frequency, such that we have an adiabatic transition.
In terms of the step-function $\Theta(x)$ (=0 or 1, respectively for $x< 0$ or $x\ge 0$), we can write the time
varying Raman frequency as $\Omega(\tau)=(45+\tau)\Theta(20-\tau)+(85-\tau)\Theta(\tau-20)$.
In these cases, the results are obtained with the same values of the nonlinear parameters, which are
related to the scattering length ratios, $\beta=\gamma=1$, implying that the inter- and intra-species scattering
lengths remain the same.

In all the cases considered in the present work, we are using exact full-numerical solutions of the coupled 
Gross-Pitaevskii (GP) formalism (\ref{eq02}),  by applying an imaginary-timer propagation method~\cite{brtka}, 
using the Crank-Nicolson algorithm, followed by real-time evolution of the soliton profiles. 
The time steps, as well as the total $x-$space interval and corresponding discretization, 
have been adapted to obtain convergent and accurate results. In most of the cases, considering our dimensionless units,  
we found the time step $\delta t=10^{-3}$ to be sufficient in the imaginary-time relaxation procedure, and $10^{-4}$ in the real-time evolution.
In this regard, we can mention that particular care has to be considered in the time-evolution of striped solitons, 
where stable solutions demand a large enough number of grid points $\delta x$ within a large $x$ interval  
(to avoid border effects). In view of that, for some results we decrease the time step to $\delta t=10^{-5}$. 
Along the text, we are providing some theoretical analysis by using variational ansatzes, 
which are verified to be more efficient in region I, where the solutions are regular solitons. For the case that rapid 
modulations are used for the Raman frequency, in region II, it was also shown to be quite useful to employ 
a multi-scale expansion for the averaged coupled system. 

\section{Time-modulated Raman frequency and internal Josephson oscillations}
An interesting case that can be explored is the influence of time-varying Raman frequency on the 
oscillations in atomic populations, which can occur between the components of the soliton solutions
(the internal Josephson effect) in the regions I and II.
The time-periodic modulation of the Raman frequency may lead to resonant responses of solitons in BEC with SOC.
This phenomenon is possible to occur for the imbalanced populations between soliton components, which are 
produced initially with different phases. To study the dynamics of this kind of process, in both the cases of region I 
($\Omega_0>k_L^2$) and region II ($\Omega_0<k_L^2$), we first  implement the Josephson oscillations for constant 
Raman frequencies $\Omega_0$, by introducing a phase between the two soliton components, before starting the time 
evolution of the coupled system. 

For the  time modulated Raman frequency,  we consider the following expression: 
\begin{equation}
\Omega(t) = \Omega_0 + \Omega_1 \cos(\omega t),  \label{eq05}
\end{equation}
where $\Omega_1$ is the amplitude, with $\omega$ the frequency of the oscillations.
Different regimes are possible in the dependence of the values of the modulating frequency $\omega$, such that we 
can have slow ($\omega \ll \Omega_0$), resonant ($\omega \sim \Omega_0$) or rapid ($\omega \gg \Omega_0$) 
modulations. 
In this section, we are mainly concerned with the intermediate regime, where we can have resonant responses,
such that we are going to assume small amplitude for the oscillations, $\Omega_1\ll \Omega_0$. 
The case of very slow frequency can be reported to the previous study presented in section II-B.
The other regime for the time-perturbed Raman, with high frequencies $\omega$ and large amplitudes $\Omega_1$, 
we are considering in  section IV.

The studies in this section are done by  a full numerical simulation, as well as by some analytical considerations 
through a variational procedure.
In order to study the interference effect on the Josephson oscillation due to a time-modulated Raman frequency,
we employ a variational approach in region I (where $k_L^2<\Omega_0$, and regular soliton solutions are 
obtained), considering the ansatz 
\begin{equation}
\left( \begin{array}{c} u \\ v \end{array} \right)=
\left( \begin{array}{c} A_1e^{-\left[{(x-x_0)^2}/{(2a^2)}- ik_1 x- {\rm i}\phi_1\right]} \\ 
A_2e^{-\left[{(x-x_0)^2}/{(2a^2)}- ik_2 x - {\rm i}\phi_2\right]} \end{array} \right)
,\label{eq06}
\end{equation}
where $A_i$, $a$ ,$x_0$, $k_i$ and $\phi_i (i=1,2)$ are time-dependent parameters, where
we have the assumption that the solitons have the same width $a$ and center-of-mass $x_0$ 
(i.e., both components overlap), which are confirmed by numerical simulations. 
By considering the Lagrangian density for Eq.~(\ref{eq02}) and the above variational ansatz,
we have
{\small
\begin{eqnarray}
{\cal L}(x,t) 
&=& \left[
\frac{i}{2}\left(u^*\frac {du}{dt}+v^*\frac {dv}{dt}\right)
+\frac{i k_L}{2}\left(u^*\frac {du}{dx}-v^*\frac {dv}{dx}\right)+cc\right]
\nonumber\\
&-&\frac{1}{2}\left|\frac {du}{dx}\right|^2-\frac{1}{2}\left|\frac {dv}{dx}\right|^2
-\Omega u^*v-\Omega v^*u\nonumber\\
&+&\frac{1}{2}|u|^4+\frac{\gamma}{2}|v|^4+{\beta}|u|^2|v|^2
\label{eq07} ,\end{eqnarray}
} with the Lagrangian given by $L=\int_{-\infty}^{\infty}dx{\cal L}(x,t)$: 
\begin{eqnarray}
L&=&-\sum_{i=1}^2N_i \left[\frac{d\phi_{i}}{d{t}} + \frac{dk_{i}}{d{t}}x_0+\frac{1}{4a^2}
+\frac{k^2_i}{2}+(-)^i k_ik_L
\right] \nonumber\\
&-&2\Omega(t) \sqrt{N_1N_2}e^{-a^2k_{-}^2}\cos(2k_{-}x_0 + \phi)\nonumber\\ 
&+&\frac{1}{2\sqrt{2\pi}\,a}(N_1^2+\gamma N_2^2+ 2\beta N_1N_2)
.\label{eq08}
\end{eqnarray}
Here, and in the following, we are using the definitions $k_{\pm}\equiv(k_1\pm k_2)/2$ and  
$\phi\equiv\phi_1-\phi_2$. The number of atoms for each component $i=1,2$ is given by 
$N_i \equiv \sqrt{\pi}A_i^2 a$, with the total number $N=N_1+N_2$ being conserved.
By assuming weak SOC parameter, $k_{-}\approx k_L$, following arguments given in 
Ref.~\cite{Malomed2017}, the corresponding Euler-Lagrange equations for the parameters   
are given by 
\begin{equation}
\left. \begin{array}{l}
\displaystyle\frac{dx_{0}}{d{t}}=k_{+},\\ \\
\displaystyle\frac{dk_{+}}{d{t}}=2 e^{-k_L^2a^2} \Omega(t)k_L \sqrt{1-Z^2}\sin(\varphi),
\\ \\
\displaystyle\frac{dZ}{d{t}}=-2e^{-k_L^2a^2} \Omega(t)\sqrt{1-Z^2}\sin(\varphi),
\\ \\
\displaystyle\frac{d\varphi}{d{t}}=2k_Lk_{+} + \Lambda Z +\frac{2e^{-k_L^2a^2} 
\Omega(t)Z}{\sqrt{1-Z^2}}\cos(\varphi) ,
\end{array}\right\}
\label{eq09}\end{equation}
\noindent where $Z\equiv (N_1-N_2)/N$, $\varphi\equiv 2k_{-}x_0 + \phi$, 
$\Lambda\equiv N (1-\beta)/(\sqrt{2\pi}a)$ and,  for simplificity, we fix $\gamma=1$.  
In case of a hyperbolic-type ansatz, the exponential factor in the above equations, derived from a 
Gaussian ansatz, is changed as $ e^{-k_L^2a^2} \to (\pi k_L/\eta)/\sinh(\pi k_L/\eta)$ (where $\eta$ is 
the soliton amplitude).  In the weak SOC limit, when $k_L \ll \eta/\pi$ or $k_L\ll 1/a$, this factor reduces to one.

This system is analogous to the one considered in \cite{Malomed2017} for a constant $\Omega=\Omega_0$,
where it was shown that in the Euler-Lagrange system for the parameters, the equation for the center-of-mass,
$x_0=(x_1+x_2)/2$, can be approximately solved as $x_0 \approx( k_L/2\Omega_0)\cos(2\Omega_0 t).$
Thus, the center-of-mass oscillations are  small, with its amplitude being of the order of $\approx k_L/2\Omega_0.$ 
By considering our numerical simulations, as discussed in more detail in the next subsection, with results
given in Figs.~\ref{fig-03} and \ref{fig-04}, for the case that $k_L=4$ and $\Omega_0 =20$, we have confirmed 
that the center-of-mass oscillations agree with the estimated value, being $\sim 0.1$, as shown in Fig.~\ref{fig-04}. 

Then, for small values of $k_L$ (as compared with $\Omega_0$), we can consider
the following coupled system:
\begin{eqnarray}
\frac{d\phi}{dt}
&=& \Lambda Z+ \frac{2{e^{-k_L^2a^2}\Omega}(t) Z}{\sqrt{1-Z^2}}\cos(\phi),\label{eq10} \\ 
\frac{dZ}{dt}&=& -2{e^{-k_L^2a^2}\Omega}(t)\sqrt{1-Z^2}\sin(\phi),
\label{eq11}\end{eqnarray}
which describes the internal Josephson oscillations of atomic populations between two pseudo-spin components.
For a constant Raman frequency, these oscillations have been studied in Ref.~\cite{Zhang2012}.
It appeared when investigating the macroscopic quantum tunneling obtained in a double-well potential having a 
barrier between the wells with constant height~\cite{Shenoy}. This was also studied in Refs.~\cite{AK,Milburn,Boukobza}
for the case that the barrier between wells have their heights oscillating in time. 

At some frequencies of the modulations, it is possible to verify parametric resonance in the Josephson oscillations. 
 In the following we will discuss the full numerical results, by considering the two possible regions
defined in Fig.~\ref{fig-01} by the relations between the Raman and SOC parameters.  The results obtained in region I, 
where we have bright type solitons, are shown to be fully compatible with a variational approach. 
 However, for the region II, where we have striped soliton solutions, the coupled system is not so amenable
to simplified variational analysis, such that we can provide only rough estimates for some limiting situations.
Therefore, in the case of time evolution for the striped solitons, with constant and time-dependent
Raman frequencies, our study relies mostly in full-numerical simulations, 
which are shown to provide convergent results with high numerical precision.

\subsection{Results for Josephson oscillations in region I}
The results in this case are for $\Omega_0>k_L^2$, considering the Raman frequency constant, as well as
time-modulated. In the case of time-modulated Raman frequency, we have also introduced a variational analysis to 
provide an estimate for the localization of the modulation frequency leading to the resonant behavior. 

\subsubsection{Constant Raman frequency: $\Omega(t)=\Omega_0$}
When $\Omega_0$ is constant, the frequency of the free oscillations is given by 
\begin{equation}
\omega_J=\sqrt{2\Omega_0(2\Omega_0 +\Lambda)} \;\;\left(= 2\Omega_0, \;{\rm for}\; \beta=1\right).
\label{eq12}
\end{equation}

\begin{figure}[tbph]
\centerline{
\includegraphics[width=8.5cm]{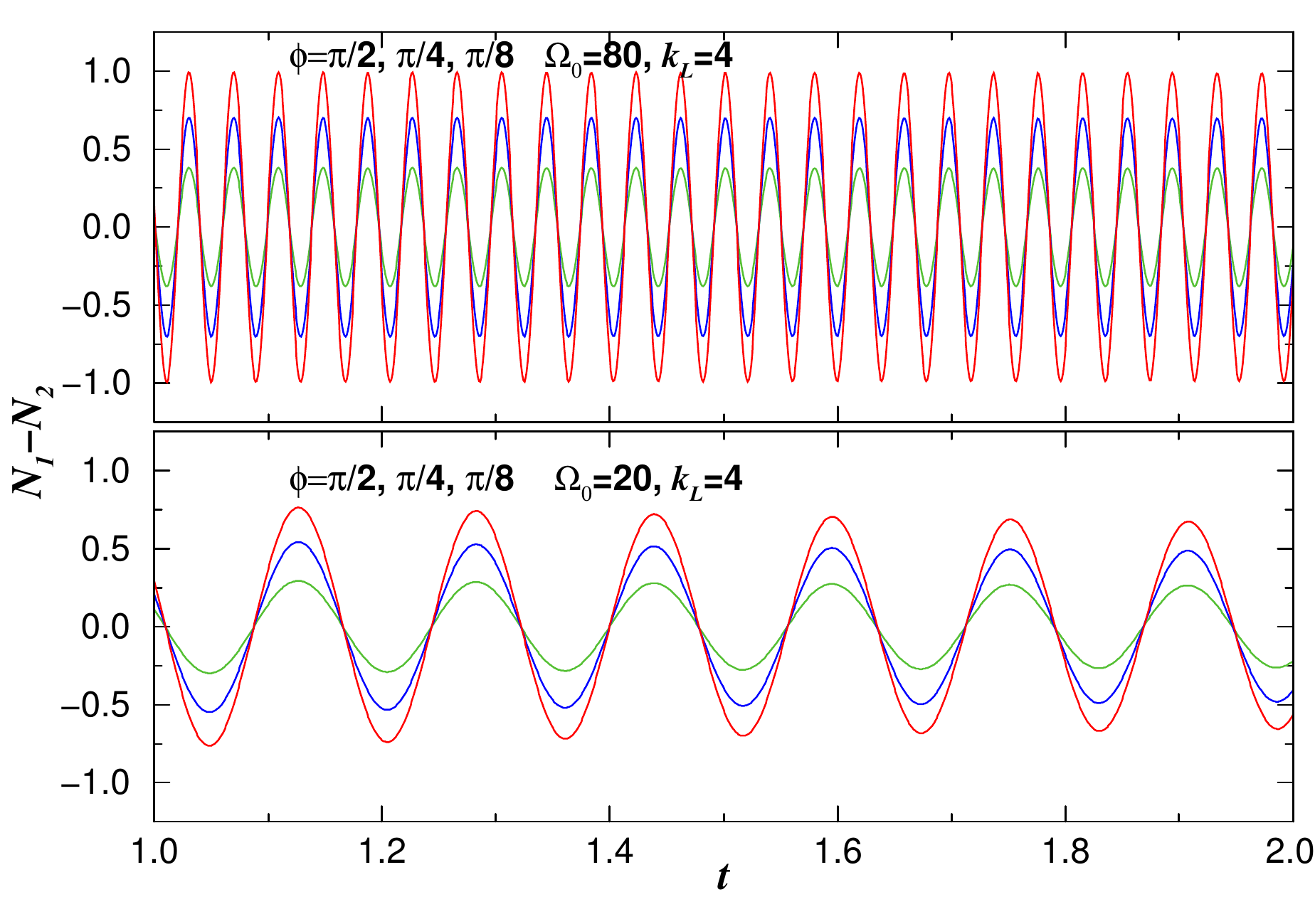}}
\caption{(color on-line)
The atom-number oscillations, with the corresponding dependence on the phase $\phi$ introduced between components
for solitons in region I, are shown for $\Omega_0=$ 80 (upper panel) and 20 (lower panel).
The given results are for $\phi= \pi/8$ (green, smaller amplitudes), 
$\pi/4$ (blue, intermediate amplitudes) and $\pi/2$ (red, larger amplitudes), 
with the spin-orbit coupling and nonlinear parameters, respectively, given by 
$k_L=4$ and $\gamma=\beta=1$. All quantities are in dimensionless units.}
\label{fig-03}
\end{figure}

\begin{figure}[tbph]
\centerline{
\includegraphics[width=9.cm]{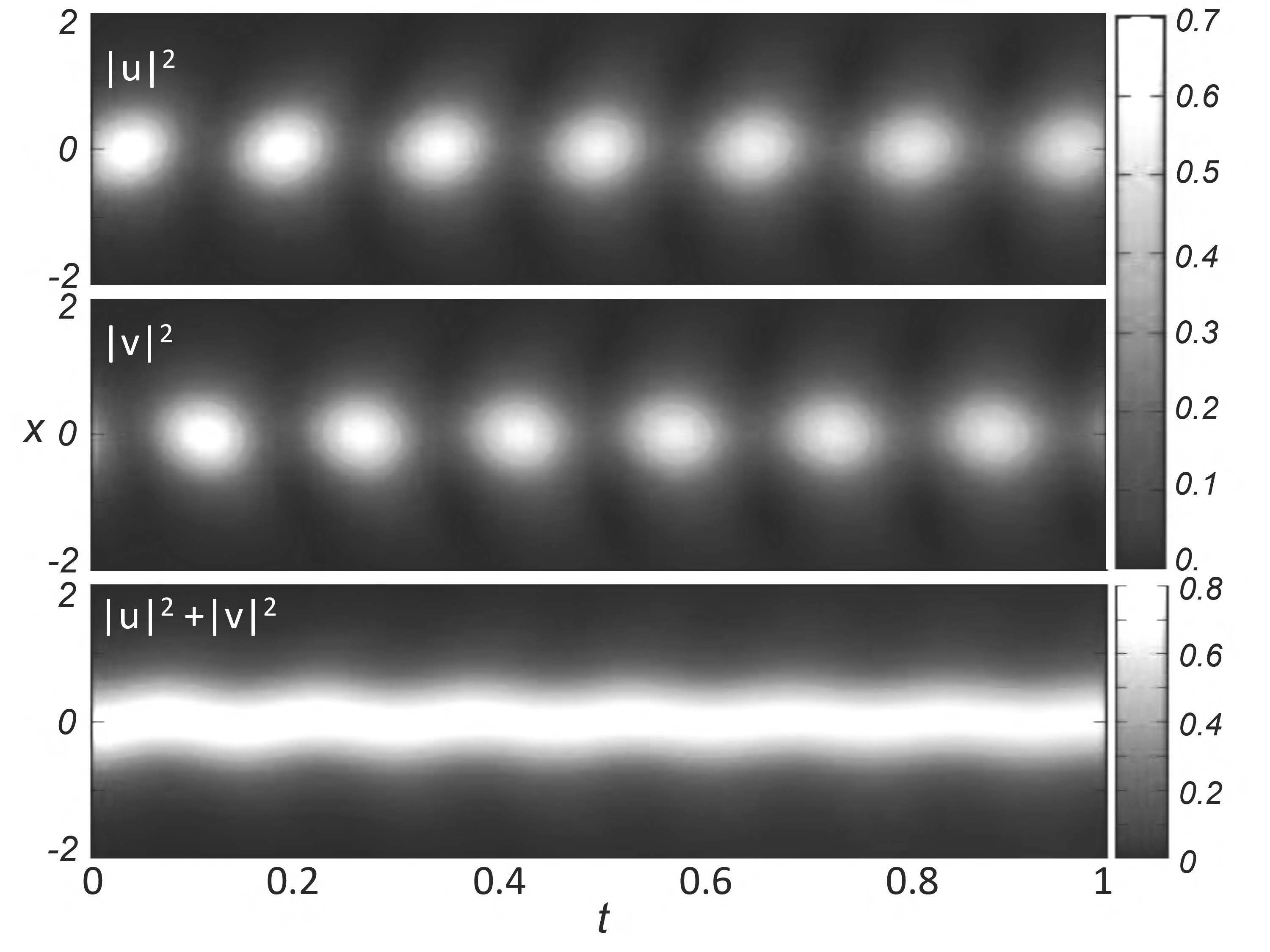}
}
\caption{Density plots for soliton profiles corresponding to the lower panel of Fig.~\ref{fig-03} ($\Omega_0=$ 20),
for $|u(x,t)|^2$ (upper panel), $|v(x,t)|^2$ (middle panel), and $|u(x,t)|^2+|v(x,t)|^2$ (lower panel), obtained at the positions $x$
in the time interval $0\le t\le 1$. The density levels are indicated in the right-hand-side. 
(all quantities are dimensionless).}
\label{fig-04}
\end{figure}

From numerical simulations for free Josephson oscillations of the full system of the GP equation, the results  
obtained in the region I ($k_L^2>\Omega_0$) are  represented in Figs.~\ref{fig-03} and \ref{fig-04}, by considering the 
spin-orbit coupling parameter $k_L=4$, with two constant Raman frequencies, given by $\Omega_0=80$ (upper panel) and 
$\Omega_0=20$ (lower panel). In Fig.~\ref{fig-04} we show the density plots, for $|u|^2$,  $|v|^2$ and $|u|^2+|v|^2$,  
corresponding to the case that $\Omega_0=20$, for the time interval $0<t<1$.

The purpose, in this case,
is to verify the dependence of the oscillating behavior on different values of the initial phase $\phi$ introduced 
between components when starting the evolution. 
By considering three values for the initial phase $\phi$, it is shown that the maximum of the periodic atom transfer 
occurs at $\phi=\pi/2$, reaching almost 100\% of atoms in the case that $\Omega_0=80$.
It is also shown that the phase $\phi$ affects only the amplitude of the oscillations, but not the frequency. 
The constant Raman parameter will determine the frequency of the 
oscillations, which is given by $\approx 2\Omega_0$, confirming the theoretical prediction (\ref{eq12}).

\subsubsection{Time-modulated Raman frequency: $\Omega(t)=\Omega_0 + \Omega_1\cos(\omega t)$}
Now, let us analyze the case when $\Omega(t)$ is modulated in time, such that 
$\Omega(t)=\Omega_0 + \Omega_1\cos(\omega t),$ in order to verify the localization of possible  
resonant behaviors. For that, we can consider two limiting conditions for the Eq.~(\ref{eq10}):
one applied for $Z\ll 1$, when $\phi(0)\approx 0$; the other for the regime of macroscopic quantum
localization, as follows.

Let us consider the  linear regime case, when $\phi(0)\approx 0$ and $Z\ll 1$.
Within these conditions in Eq.~(\ref{eq10}), 
using the second derivative for $Z$, we obtain a modified Mathieu differential equation, with the 
main term in $Z$ oscillating with a frequency $\omega_0$, where 
$\omega_0^2 \equiv 2\Omega_0(2\Omega_0+\Lambda)+(2\Omega_1)^2.$
For that, when considering small values of $\Lambda$ and  $\Omega_1$, the standard analysis as given 
in Ref.~\cite{LL} can be applied, which leads to  
a resonance at $\omega=2\Omega_0$ (in case $\beta=1$, so $ \Lambda=0$). In this case,  
parametric resonances are also expected to occur for  $\omega=2\omega_J$, 
where $\omega_J$ is the frequency of free Josephson oscillations given in Eq.~(\ref{eq12}).

\begin{figure}[tbph]
\centerline{
\includegraphics[width=9cm]{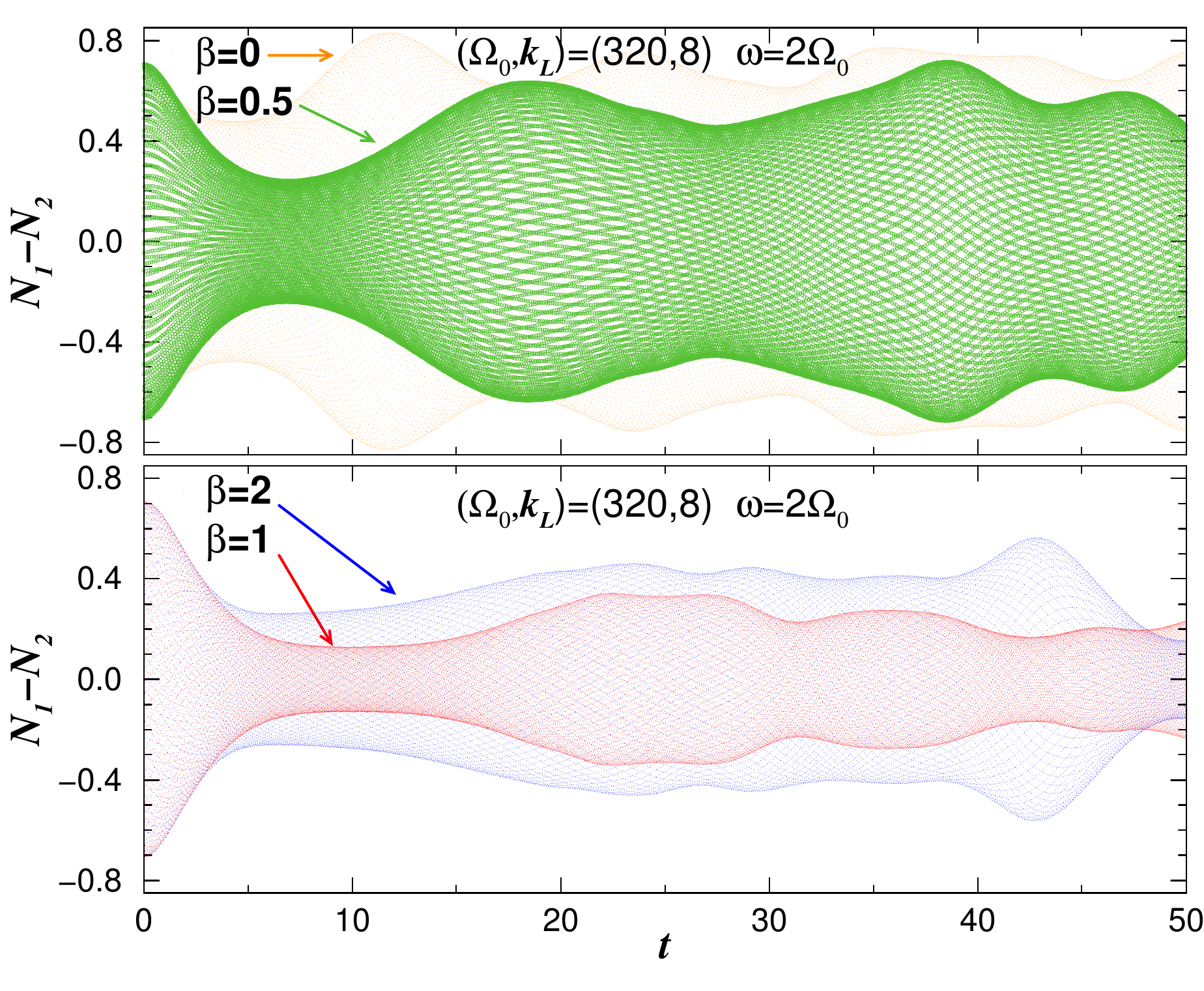}}
\caption{(color on-line) Resonant interference patterns verified for $\omega =2\Omega_0$ are shown in the atom-number 
oscillations for the case that $\Omega_0=320$, $\Omega_1=0.1 \Omega_0$, $k_L=8$ and $\gamma=1$, considering
different values of the parameter $\beta$ (as indicated by the corresponding arrows). We have
$\beta=$0 and 0.5 in the upper panel; and $\beta=$1 and 2 in the lower panel.
The shadowing areas, in each case, 
are representing the range of the oscillations in the real-time propagation of the two-components. In all these cases, the time
evolution was performed with $\delta t=10^{-4}$, with a starting phase $\pi/4$ introduced between the two components. 
All quantities are in dimensionless units.
}
\label{fig-05}
\end{figure}

\begin{figure*}[tbph]
\centerline{
\includegraphics[width=18cm]{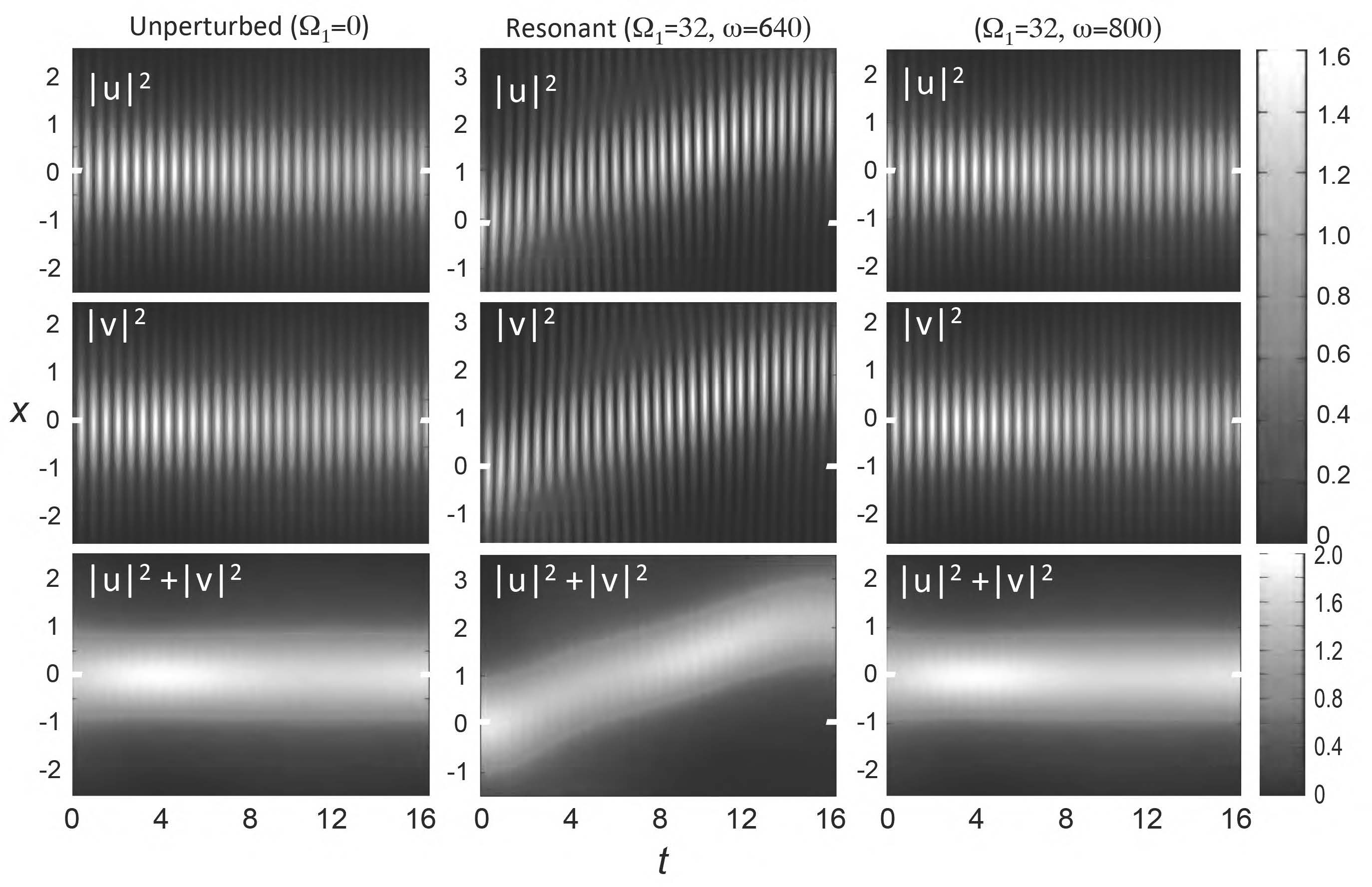}}
\vspace{-0.5cm}
\caption{
Density plots for soliton profiles corresponding to the case with $\beta=0$ shown in Fig.~\ref{fig-05}
($\Omega_0=$ 320), for $|u(x,t)|^2$ (upper panels), $|v(x,t)|^2$ (middle panels), and 
$|u(x,t)|^2+|v(x,t)|^2$ (lower panels), obtained at the positions $x$ in the time interval $0\le t\le 16$.
In the left frames we have the non-perturbed oscillations ($\Omega_1=0$);
in the middle frames, the oscillations at resonant position ($\Omega_1=0.1\Omega_0=32$, with 
$\omega=2\Omega_0$);
and, in the right frames, the non-resonant perturbed case, with $\omega=2.5\Omega_0$. 
All quantities are in dimensionless units.
}
\label{fig-06}
\end{figure*}

\begin{figure}[tbph]
\centerline{
\includegraphics[width=8.8cm]{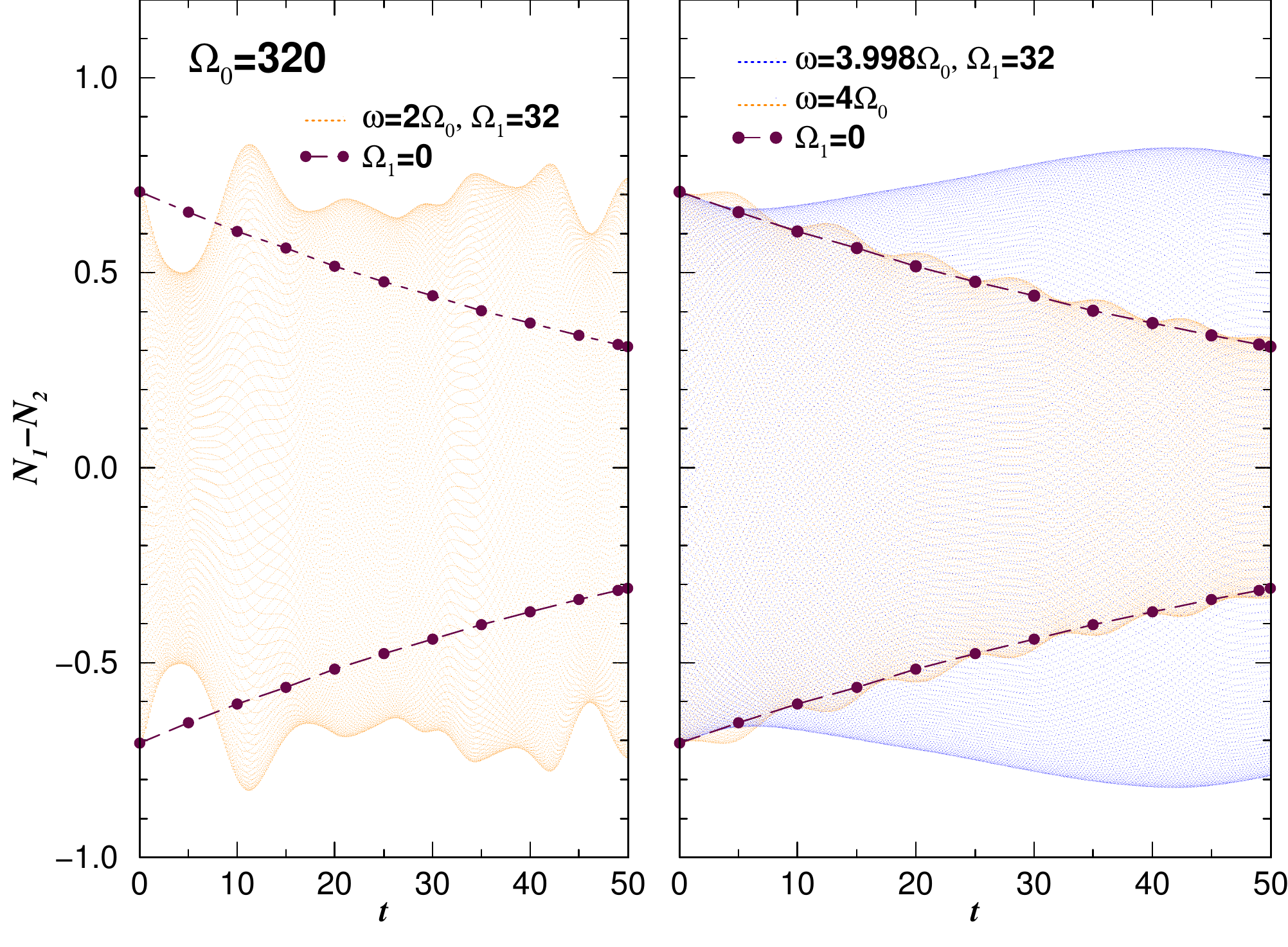}}
\caption{(color on-line) 
Parametric resonant behaviors verified in the atom-number oscillations $N_1-N_2$ during the time-propagation of a 
two-component soliton, with $\Omega_0=320$, $k_L=8$, $\Omega_1=0.1\Omega_0$, $\beta=0$ and 
$\gamma=1$. In both panels, we also indicate with dashed-bullet lines the extremes of the oscillations for the
case $\Omega_1=0$, for comparison.
In the left panel, the oscillations at resonant position, $\omega=2\Omega_0$, are within the shadowing region.
In the right panel, we show the results for two frequencies ($\omega=4\Omega_0$ and $\omega=3.998\Omega_0$)
close to the region where parametric resonance are expected. The shadowing region with oscillations close
to the non-perturbed case is for $\omega=4\Omega_0$. 
In all these cases, for the time interval shown, the time step was $\delta t=10^{-5}$, with a starting phase $\pi/4$ 
introduced between the two components.  All quantities are in dimensionless units.
}
\label{fig-07}
\end{figure}

\begin{figure}[tbph]
\centerline{
\includegraphics[width=9cm]{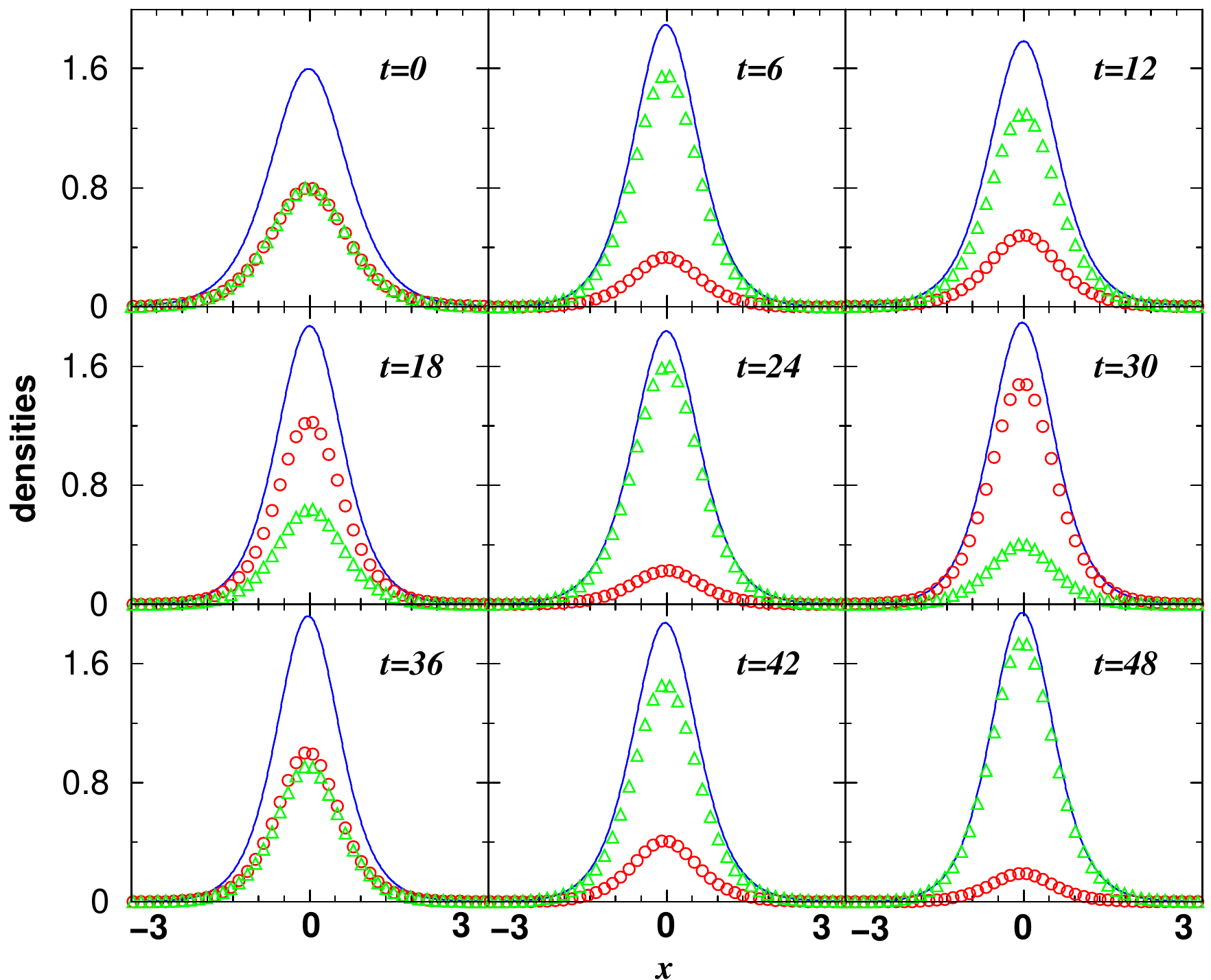}}
\caption{(color on-line) Density profiles, $|\psi|^2$ (solid-blue), $|u|^2$ (circled-red) and $|v|^2$ (triangle-green) 
for different time positions in the evolution of the coupled soliton, considering the case shown in the
right panel of Fig.~\ref{fig-07}, where $\omega=3.998\Omega_0$.
The parameters are $k_L=8$, $\Omega_0=320$, $\Omega_1=0.1 \Omega_0$, $\beta=0$ and $\gamma=1$, with a  
starting phase $\pi/4$. All quantities are in dimensionless units.
}
\label{fig-08}
\end{figure}

Results of our investigations on resonant interferences which can occur in region I are being presented in 
Figs.~\ref{fig-05} to \ref{fig-08}. They are obtained from numerical simulations of the full coupled system (\ref{eq02}), with the Raman 
and SOC parameters $\Omega_0=320$ and $k_L=8$, respectively. For the time modulation of the Raman parameter, given by Eq.~(\ref{eq05}), we assume the amplitude given by $\Omega_1 = 0.1\Omega_0$. 
The choice of these parameters in region I are to distinguish more clearly the manifestation of resonant interferences in the 
Josephson oscillations between the atom numbers of the two components, $Z=N_1-N_2$, $(N=1)$.
We found also illustrative to provide some density plots, for the components and total profiles, corresponding to the results 
presented in Fig.~\ref{fig-05}. Therefore, in Fig.~\ref{fig-06} we are showing the case where we have $\beta=0$ in Fig.~\ref{fig-05}. 
Here, we should remark that, the perturbed results obtained when we are not close to the resonant interference regions are 
shown to be almost identical to the unperturbed results (as observed by comparing the first column of panels of Fig.~\ref{fig-06} 
with the third column). Indeed, quite small center-of-mass oscillations are verified near the initial localization.
Another point that can be observed from these results is that the center of mass of the soliton is strongly affected by the 
resonant behavior, such that, it can be verified in the central panels of Fig.~\ref{fig-06} that the central position is moving 
from $x=0$ at $t=0$ to $x\approx 2$ at $t=16$. There is no change observed in the center of mass for the other cases, 
outside the resonant region.  
The numerical simulations of the variational system (\ref{eq08}) confirms this behavior of the motion of the center 
of mass for the solitonic components and the oscillations of the atomic imbalance at the resonance.

We should comment that, for the values of the frequency $\omega$, the resonant position (``window") is quite sharp,
verified in our numerical simulation, such that the resonant perturbations are being confirmed only for $\omega$ very
close to  2 and 4 $\Omega_0$, which makes the simulations quite time demanding. As shown in Fig.~\ref{fig-05} and
in the left panel of Fig.~\ref{fig-07}, for a large time interval going up to $t=50$, one of the resonant position are detected 
for $\omega=(2\pm0.001)\Omega_0$. By a slight larger deviation of this frequency the results for the oscillations are
about the same as given for the non-perturbed case ($\Omega_1=0$) shown in the left panel of Fig.~\ref{fig-07}.
The other resonant position, as shown in the right panel of Fig.~\ref{fig-07}, for $\beta=0$, is found in an even smaller 
range of $\omega$, given by $\omega=3.998\Omega_0$, with fluctuation start appearing when we have $\omega$
exact $4\Omega_0$. 
In Fig.~\ref{fig-05} we are also showing how the resonant behaviors are affected by changes in the nonlinear 
parameter $\beta$. As shown it is enhanced in case that $\beta=0$, with the variation $N_1-N_2$ having peaks with 
maxima close to 0.9.
The case of $\beta=0$, for $\omega=2\Omega_0$, is also presented in the left frame of Fig.~\ref{fig-07},
for comparison with the unperturbed results of the Josephson oscillations.  

With Fig.~\ref{fig-08}, we conclude the analysis of the results shown in Figs.~\ref{fig-05} Figs.~\ref{fig-06} and \ref{fig-07}, 
by presenting the behaviors of density profiles (total and for each component) in two-dimensional plots,  for different time 
positions of the evolution.

\subsection{Results for Josephson oscillations in region II}
Now, let us consider the Josephson oscillations between components of the striped soliton solutions, corresponding to 
the region II in the dispersion relation, which are given by $\Omega_0<k_L^2$.  We perform this study by considering 
full numerical simulations of the corresponding GP coupled equations. First, we provide some results obtained for 
Josephson oscillations in the case that we have constant Raman frequency. Next, we consider the more general
case, where the Raman frequency is modulated in time, and we can have resonant results at some particular
values of the modulating frequency $\omega$.

We start the study of this section by considering a variational analysis, where we need to introduce the momentum 
$k_0=\pm k_L\sqrt{1-\Omega_0^2/k_L^4}$,
which provides the momentum position of the minima shown in the lower panel of Fig.~\ref{fig-01}. 
By observing that the coupled equations for the imbalanced populations and relative phase are not easy to be derived in 
explicit form, in a more general case, let us assume that the tunneling between components occurs for the same sign 
of $k_0$.  The time modulations for $\Omega(t)$ are not inducing transitions between oppositely propagating modes 
with $\pm k_0$. To have such transitions, so called momentum Josephson oscillations, we need parameters with 
periodic modulation in space~\cite{stripeJoseph,2016-Luo}. 
Therefore,  by assuming for the components the same ansatz as given in Eq.~(\ref{eq06}), but with 
$k_{i=1,2}=k_0$ and considering the center fixed at $x=0$, we arrive to the same coupled expressions  
(\ref{eq10}) and (\ref{eq11}), except that the equation for $d\phi/dt$ contains an additional term $2k_L k_0$.
By linearizing the system relative to $Z$, we obtain
\begin{eqnarray}
\frac{d\phi}{dt} &=& 2k_L k_0 +\left[\Lambda + 2\Omega(t)\right]Z\cos(\phi),
\label{eq13} \\
\frac{dZ}{dt} &=& -2\Omega(t)\sin(\phi) .
\label{eq14}\end{eqnarray}
For a  constant $\Omega$ and with $2k_Lk_0 \gg \Lambda + 2\Omega_0$, we have
\begin{equation} 
\phi \approx \phi_0 + (2k_Lk_0)t,
\label{eq15}\end{equation}
implying that the population imbalance $Z$ is oscillating with the frequency 
$\omega_{str} \approx 2k_L k_0$.
Therefore, we should expect a different behavior of the results, in comparison with the region I, in the initial 
stage (defined by $k_L k_0$).  For larger time of the propagation, the frequency for the oscillations should 
approach the same ones as verified for the region I.

\subsubsection{Constant Raman frequency: $\Omega(t)=\Omega_0$}
\begin{figure}[tbph]
\centerline{
\includegraphics[width=8cm]{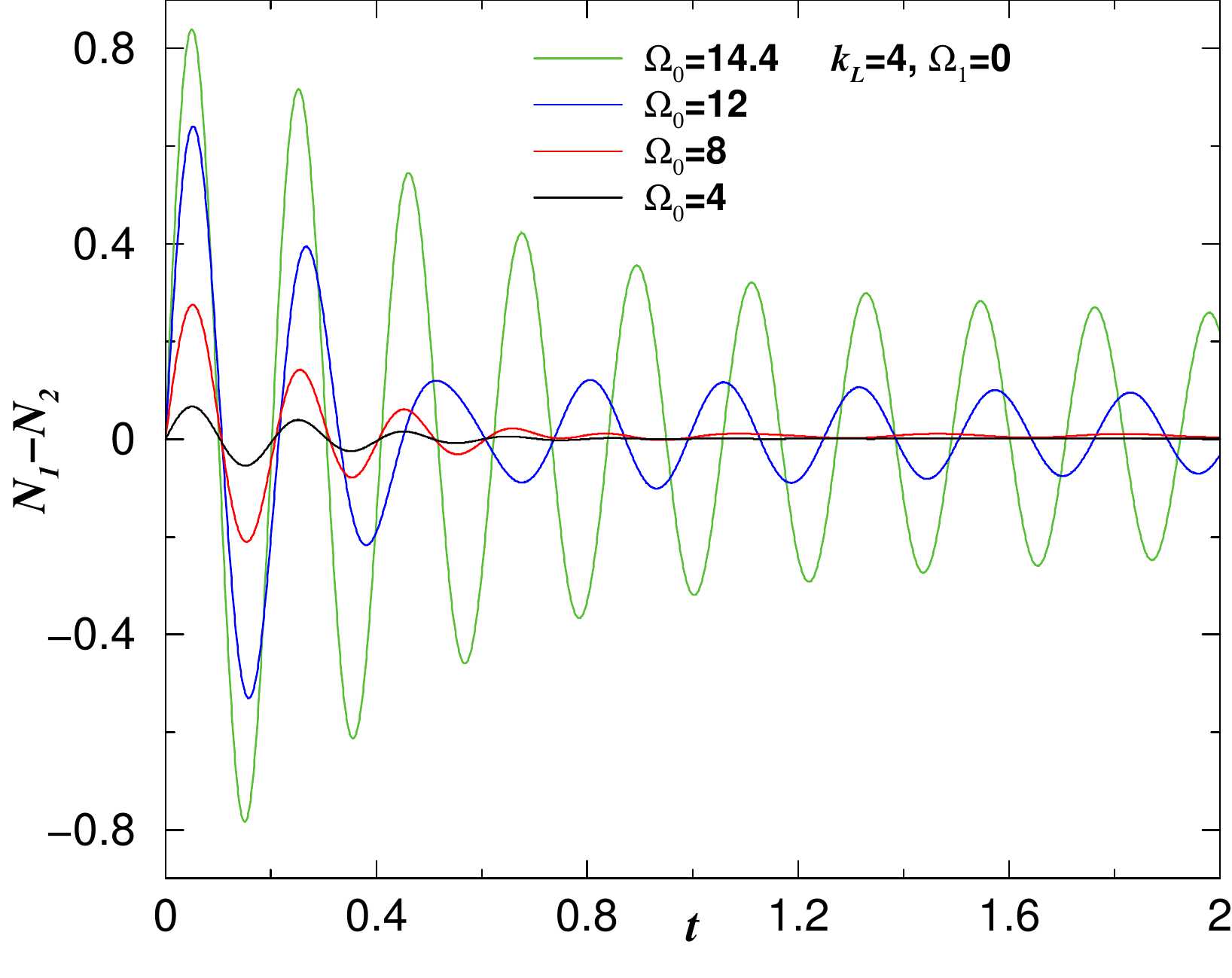}}
\caption{(color on-line) Atom-number oscillations between components are shown as functions of time, 
for $k_L=4$ in the region II, where larger initial amplitudes correspond to 
larger values of $\Omega_0$ (indicated inside the panel). 
As verified, the amplitude of the oscillations decay faster for smaller values of $\Omega_0$. 
The initial phase is $\phi=\pi/2$ (to enhance the amplitude of the oscillations), with $\beta=\gamma=1$. 
All quantities are dimensionless.
}
\label{fig-09}
\end{figure}

\begin{figure}[tbph]
\centerline{
\includegraphics[width=8.5cm]{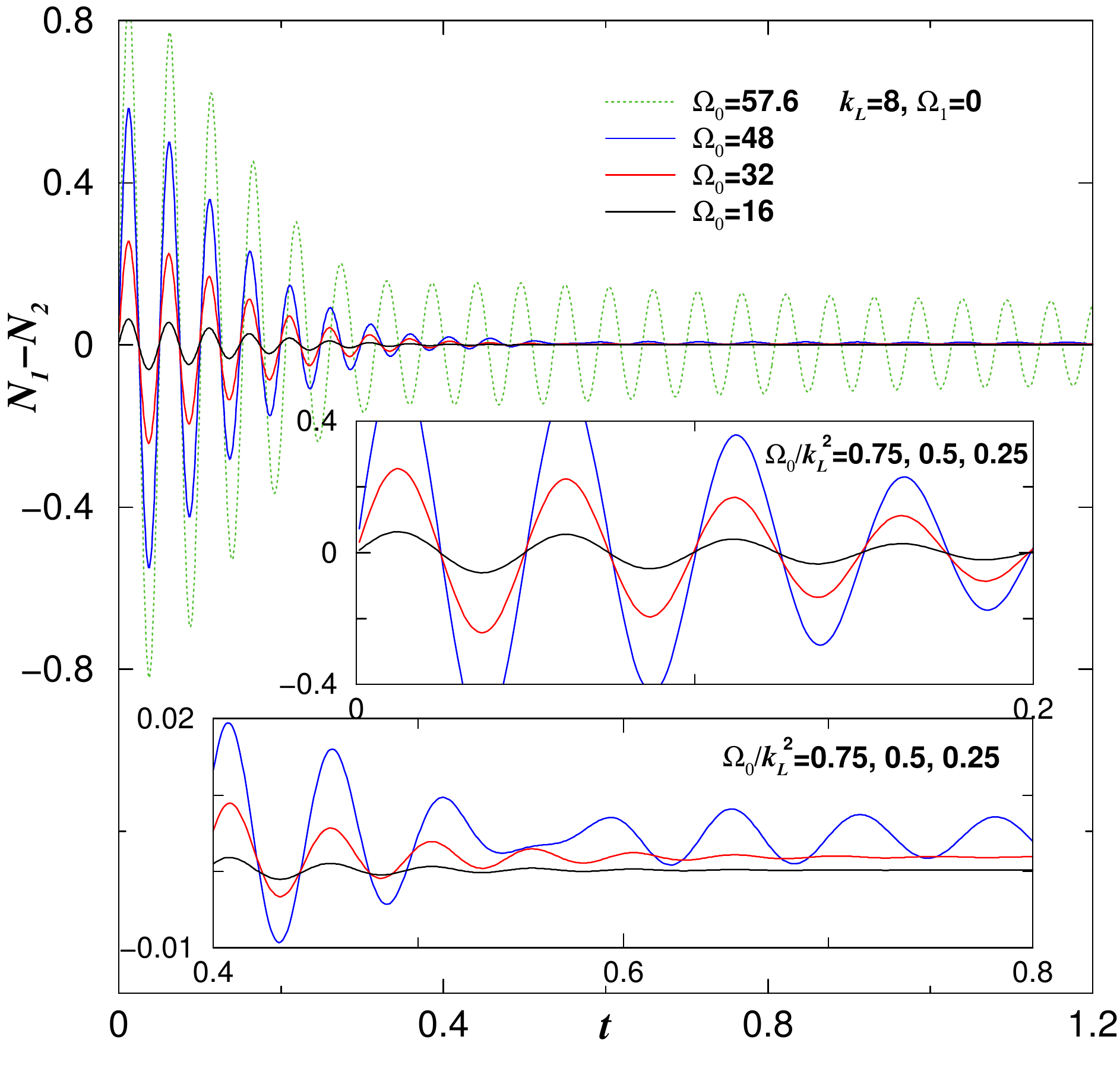}}
\caption{(color on-line) Atom-number oscillations between components are shown as functions of time,
for $k_L=8$ in the region II,  for few values of $\Omega_0$ (indicated inside the panels). 
As in Fig.~\ref{fig-09}, larger initial amplitudes correspond to larger values of $Omega_0$.
The two inset panels are given in appropriate scales to clarify the change in the 
oscillating behavior in two time intervals. 
As in Fig.~\ref{fig-09}, here the initial phase was fixed at $\phi=\pi/2$, with $\beta=\gamma=1$. 
All quantities are dimensionless.}
\label{fig-10}
\end{figure}

For the numerical simulation of the Josephson oscillations obtained in region II, we first select some results
obtained for constant values of $\Omega_0$, which are given in Figs.~\ref{fig-09} and \ref{fig-10}, by considering
an initial phase difference between components given by $\pi/2$.
In these cases, by considering $k_L=4$ (Fig.~\ref{fig-09}) and $k_L=8$ (Fig.~\ref{fig-10}), with several values of 
$\Omega_0<k_L^2$, we can verify clearly that we have an initial stage of the oscillations, where the frequencies 
are not depending on $\Omega_0$, but only on the values of $k_L$, being $\omega\sim 10\pi$ for $k_L=4$ 
and $\omega\sim 40\pi$ for $k_L=8$. The values of $\Omega_0$ affects only the 
amplitude of the initial oscillations.  However, for larger times, the behavior of the frequencies are similar as in 
region I. Another observation from these results is that, for long-time interval
the Josephson oscillations are being damped as we increase the difference $k_L^2-\Omega_0$. In Fig.~\ref{fig-10},
we present two inset panels from where we can verify the initial and intermediate oscillation patterns.
In the inset with $t\le 0.2$, just after starting the evolution, the frequency is about the same for all the three cases, 
$\omega\sim 40\pi$, not depending on $\Omega_0$. In the other inset, for  $0.2\le t\le 0.4$, after a transient 
time interval, the frequency change to $2\Omega_0$ (as in case of region I). 


\subsubsection{Time-modulated Raman frequency: $\Omega(t)=\Omega_0 + \Omega_1\cos(\omega t)$}

When studying the phase-dependence of the atom-number oscillations for striped soliton solutions, we first observe that
the amplitude of the oscillations depends on the initial phase difference, as already verified in the case that we have
constant Raman frequency parameter, given by $\Omega_0$.
Therefore, before considering the case where we have the Raman frequency perturbed in time, we have studied the 
phase-dependence of the atom-number oscillations for striped soliton solutions during time evolution. In this numerical
study, we have verified that for arbitrary initial fixed phase $\phi$ (from 0.01 to $\pi/2$) introduced between components, 
only the amplitude of the oscillations are being affected, which are being verified by the transient time just after starting 
the evolution of the solutions. The frequency of the oscillations 
does not depend on the strength of the Raman frequency $\Omega_0$, at least during the transient time 
till the oscillations become stable. In a longer time interval, after the transient time, the frequency of the
oscillations will correspond to the Raman frequency, being given by $2\Omega_0$, as discussed for 
the case of regular soliton solutions. 
As the initial phase between the components can be arbitrary and will not affect the natural frequency of
the oscillations between $N_1$ and $N_2$, when studying time-perturbed Raman frequency, in general 
we choose this phase to be $\pi/4$, such that the amplitude of the natural oscillations is not too large,
as well as not too small. 

\begin{figure}[tbph]
\centerline{
\includegraphics[width=8.5cm]{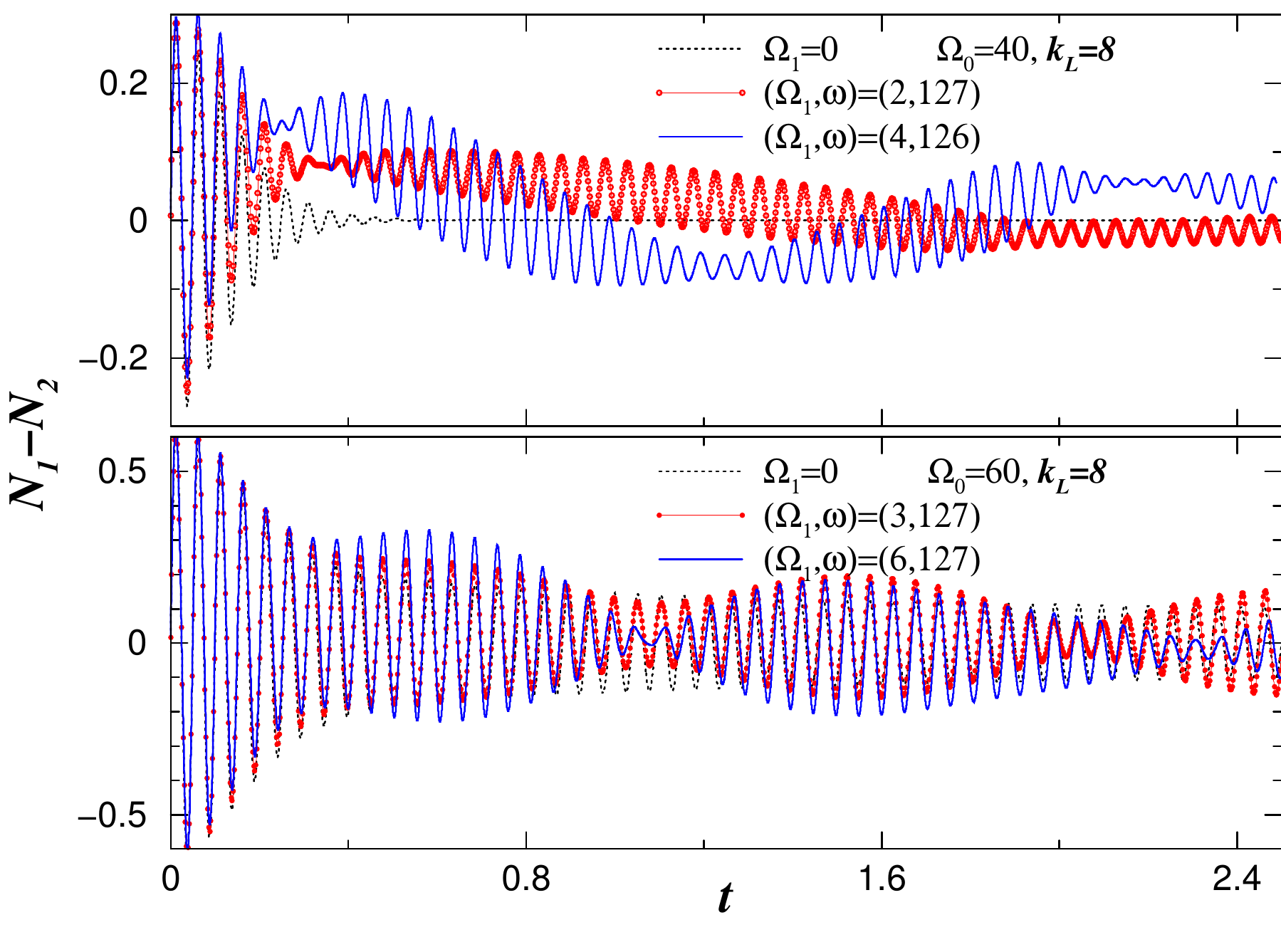}}
\caption{(color on-line) Resonant interferences in the atom-number oscillations are shown for 
striped solitons, for $k_L=8$, $\beta=\gamma=1$ and initial phase  $\pi/4$.
In the upper panel, we have $\Omega_0=40$, with  $\Omega_1=$2 (red-with-circles line) and 4 (blue-solid line) 
being compared with the unperturbed case $\Omega_1=$0 (black-dotted line). In the lower panel, for
$\Omega_0=60$ more close to $k_L^2$, we have $\Omega_1=$3 (red-with-circles line) and 6 (blue-solid line),
being compared with the case $\Omega_1=$0 (black-dotted line).  In both the cases, the resonant interferences 
are verified for $\omega\sim 40\pi$. All quantities are in dimensionless units. 
}
\label{fig-11}
\end{figure}

\begin{figure}[tbph]
\centerline{
\includegraphics[width=8.5cm]{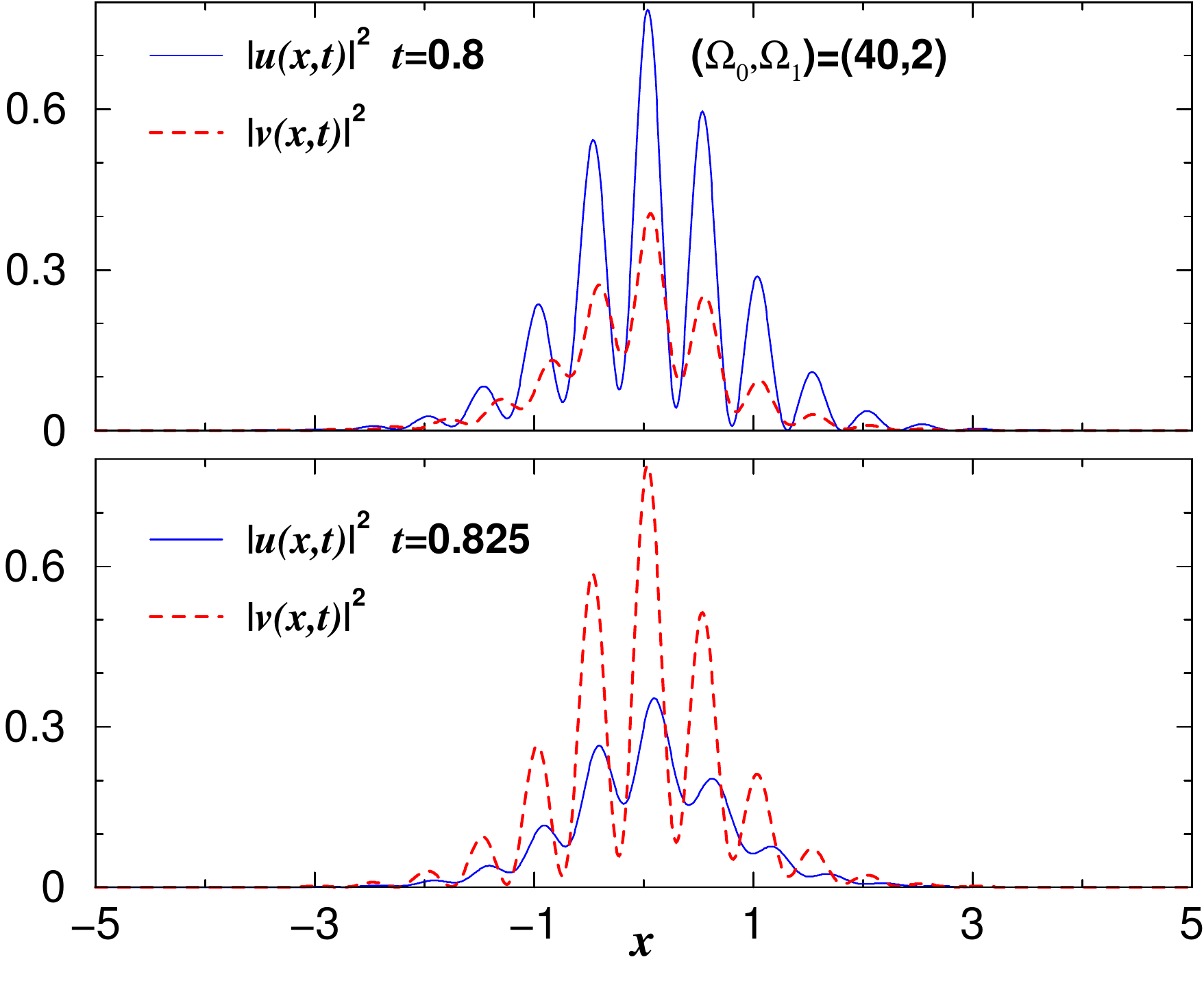}}
\caption{(color on-line) 
Dynamics of the density oscillations between components for striped-solitons are represented 
by two static panels: In the upper panel, the densities $|u(x,t)|^2$ and $|v(x,t)|^2$ are given for 
a fixed instant of time $t=0.8$, with the lower panel given for $t=0.825$, where the period of 
oscillations is about 0.05 (as verified from Fig.~\ref{fig-11}). 
The corresponding values of $\Omega_0$ and $\Omega_1$ are indicated inside the panels.
The other parameters are such that  $\beta=\gamma=1$, $k_L=8$. All quantities are in dimensionless units.
}
\label{fig-12}
\end{figure}

\begin{figure}[tbph]
\centerline{
\includegraphics[width=8.5cm]{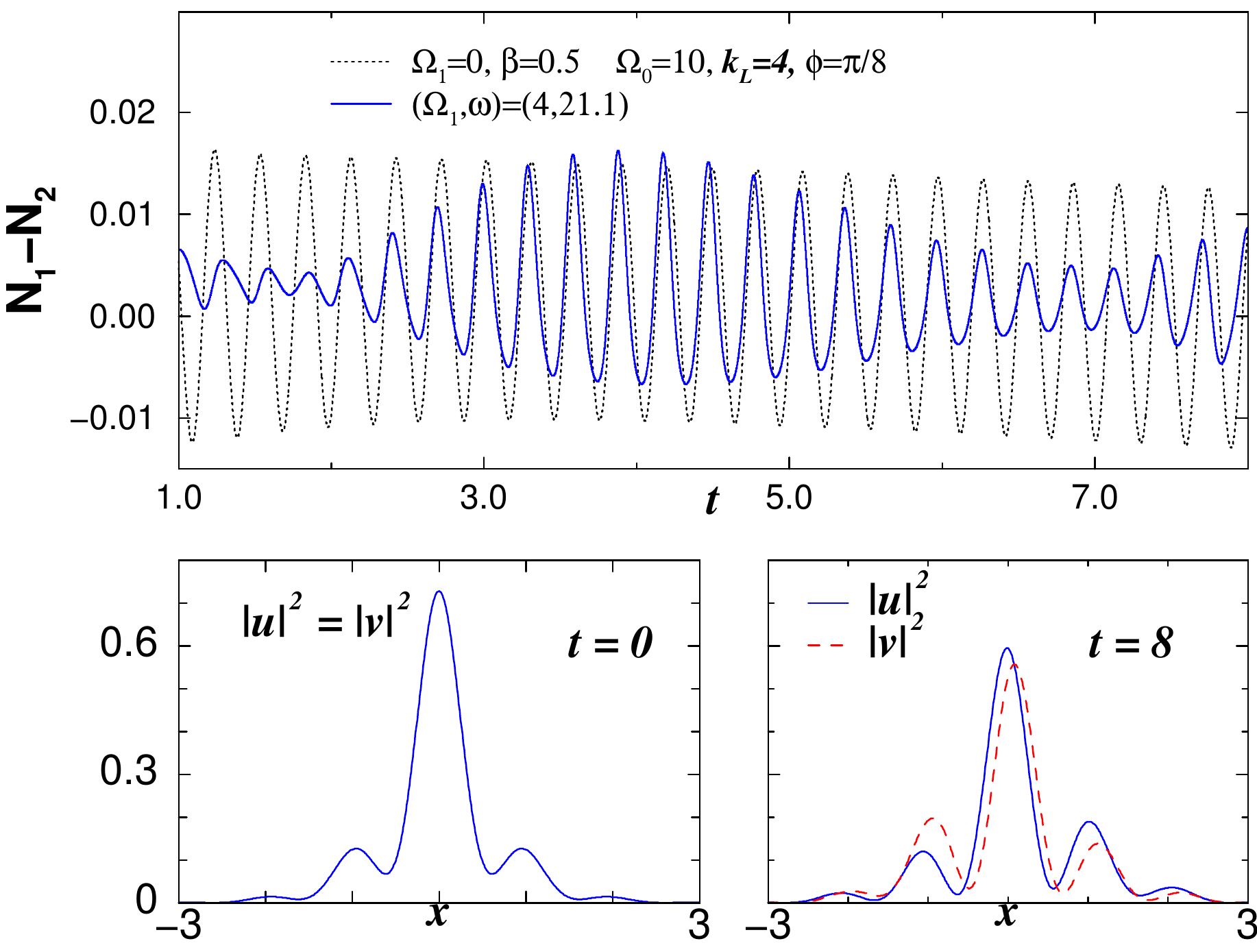}}
\caption{(color on-line) Atom-number oscillations, $N_1-N_2$ (upper panel), for striped solitons obtained in region II, 
for $k_L=4$, $\Omega_0=10$, with corresponding resonant behavior that occurs at $\omega=21.1$.
In the two lower panels we show the densities for the two components at $t=0$ and $t=8$, where the 
second case shows the effect of the time perturbation that was introduced. 
The other parameters are $\gamma=1$, $\beta=0.5$, and the initial phase $\phi=\pi/8$.
All quantities are dimensionless.
}
\label{fig-13}
\end{figure}

In the case of periodic modulations of $\Omega(t)$ in the region II ($\Omega_0<k_L^2$) the results of our 
numerical simulations to identify parametric resonances is first being exemplified by the Figs.~\ref{fig-11}
and \ref{fig-12}. In Fig.~\ref{fig-11}, for $k_L=8$, we present two panels considering $\Omega_0=$ 40 (upper) and 
60 (lower). In both we are plotting the perturbed case considering the amplitude of the oscillations given by 
$\Omega_1=0.05\Omega_0$ and $0.1\Omega_0$, with the frequency $\omega\approx 2\pi\times$ the linear 
frequency verified in the transient time interval (about 126$\sim$127). The non-perturbed case ($\Omega_1=0$)
is also shown for comparison, in both the cases.
We should emphasize that in general the Josephson oscillating behavior is about the same as it happens for the 
unperturbed case $\Omega_1=0$, except close to the specific values for $\omega$ and $\Omega_1$ where 
resonant interference behaviors are detected.
With Fig.~\ref{fig-12}, for a time interval of half-period of the Josephson oscillations,  
we are representing the profiles of the two component densities ($|u(x,t)|^2$ and $|v(x,t)|^2$),
for the case shown in the upper panel of Fig.~\ref{fig-11} with $\Omega_0=40$ and $\Omega_1=2$ (the other 
parameters are the same). The oscillation dynamics is represented  in two panels, given for $t=0.8$ (upper panel) 
and $t=0.825$ (lower panel), considering that a complete period is close to $\approx 0.05$.
The panels are indicating (through the corresponding densities) the atom-number oscillation between the 
components

For a long-time interval, resonant behaviors are expected to occur when considering cases where the natural 
frequency of the oscillations are still surviving in the unperturbed case. For that, in Fig.~\ref{fig-13}, we are showing
results for a simulation with $k_L=4$ and $\Omega_0=10$, with $\beta=0.5$ and initial phase $\pi/8$. 
In this case, by taking $\Omega_1=4$, we can observe a resonant interference that occurs for 
$\omega_1\sim2\Omega_0$. For this case, the striped soliton profiles of both components are also being represented in 
the two lower panels of the figure. In the left panel we have them at $t=0$, and in the right panel for
$t=8$.

\begin{figure}[tbph]
\centerline{
\includegraphics[width=9cm]{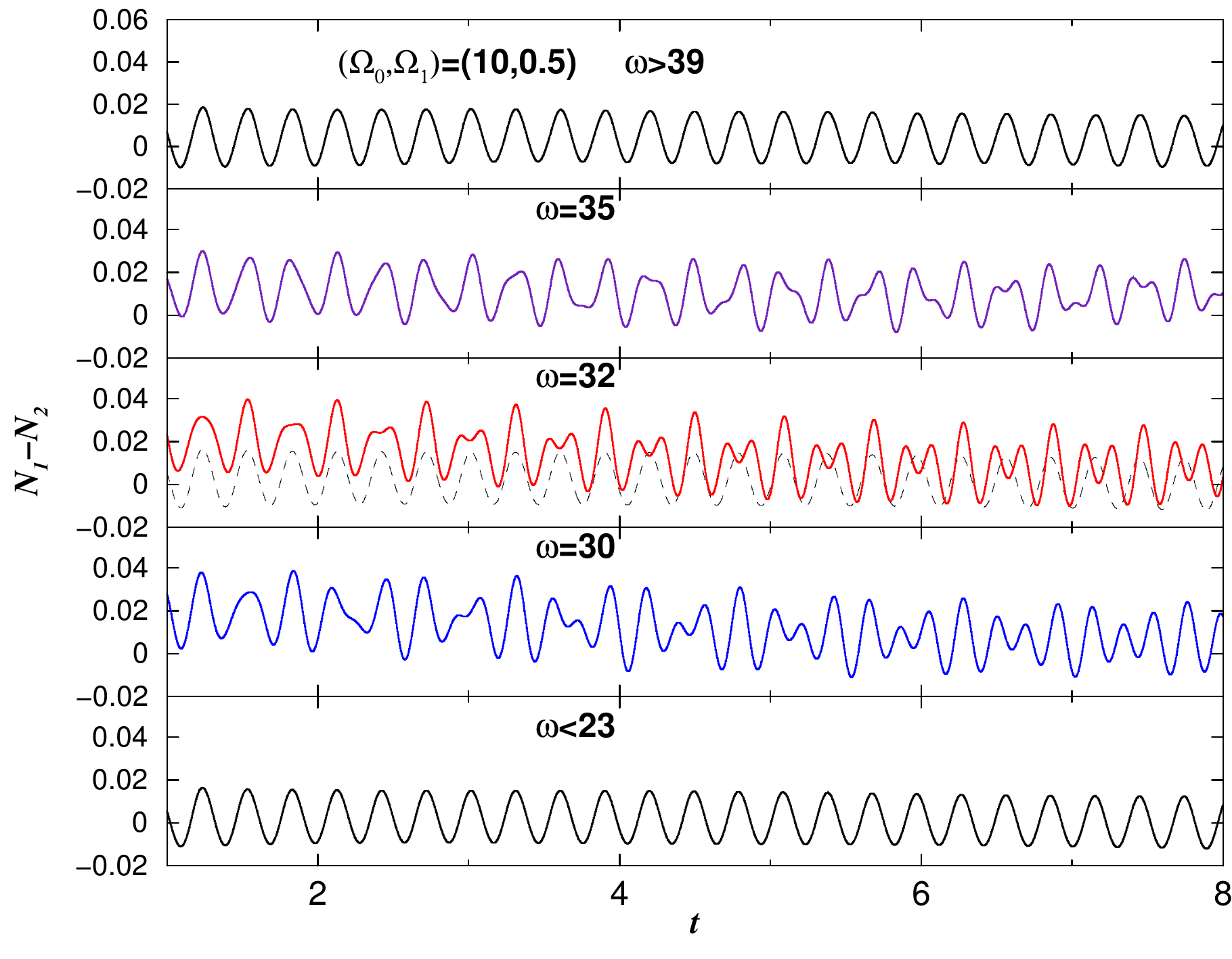}}
\caption{(color on-line) Atom-number oscillations, $N_1-N_2$, for striped solitons in region II, 
for $k_L=4$ and $\Omega_0=10$, with fixed very small amplitude $\Omega_1=0.5$ and initial 
phase between components given by $\phi=\pi/8$. The frequency $\omega$ of the time-perturbed
Raman is being varied within the region where resonant interferences are verified. For $\omega=32$,
the unperturbed case is also shown with dashed-line.
As in Fig.~\ref{fig-13}, the nonlinear parameters are $\gamma=1$ and $\beta=0.5$.
All quantities are in dimensionless units.
}
\label{fig-14}
\end{figure}

\begin{figure}[tbph]
\centerline{
\includegraphics[width=9.cm]{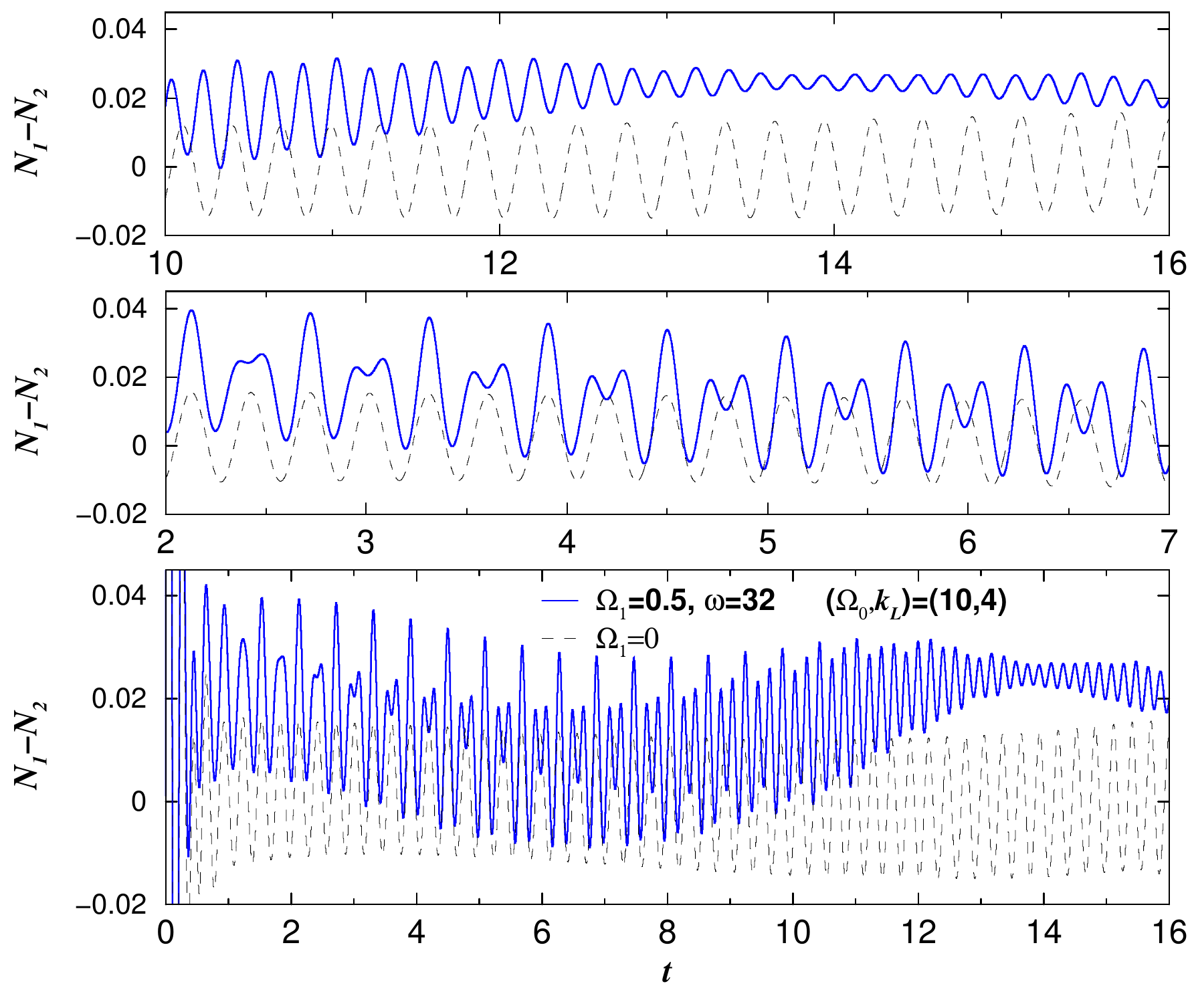}}
\caption{(color on-line) 
The case with $\omega=32$ (solid lines) was selected from Fig.~\ref{fig-14}, including 
the unperturbed case (black-dashed line). In the lower panel we have a larger time interval ($0\le t\le16$), 
with the middle and upper panels having smaller intervals given by $2\le t\le7$ and $10\le t\le16$, respectively.
All quantities are dimensionless.
}
\label{fig-15}
\end{figure}

\begin{figure}[tbph]
\centerline{
\includegraphics[width=10cm]{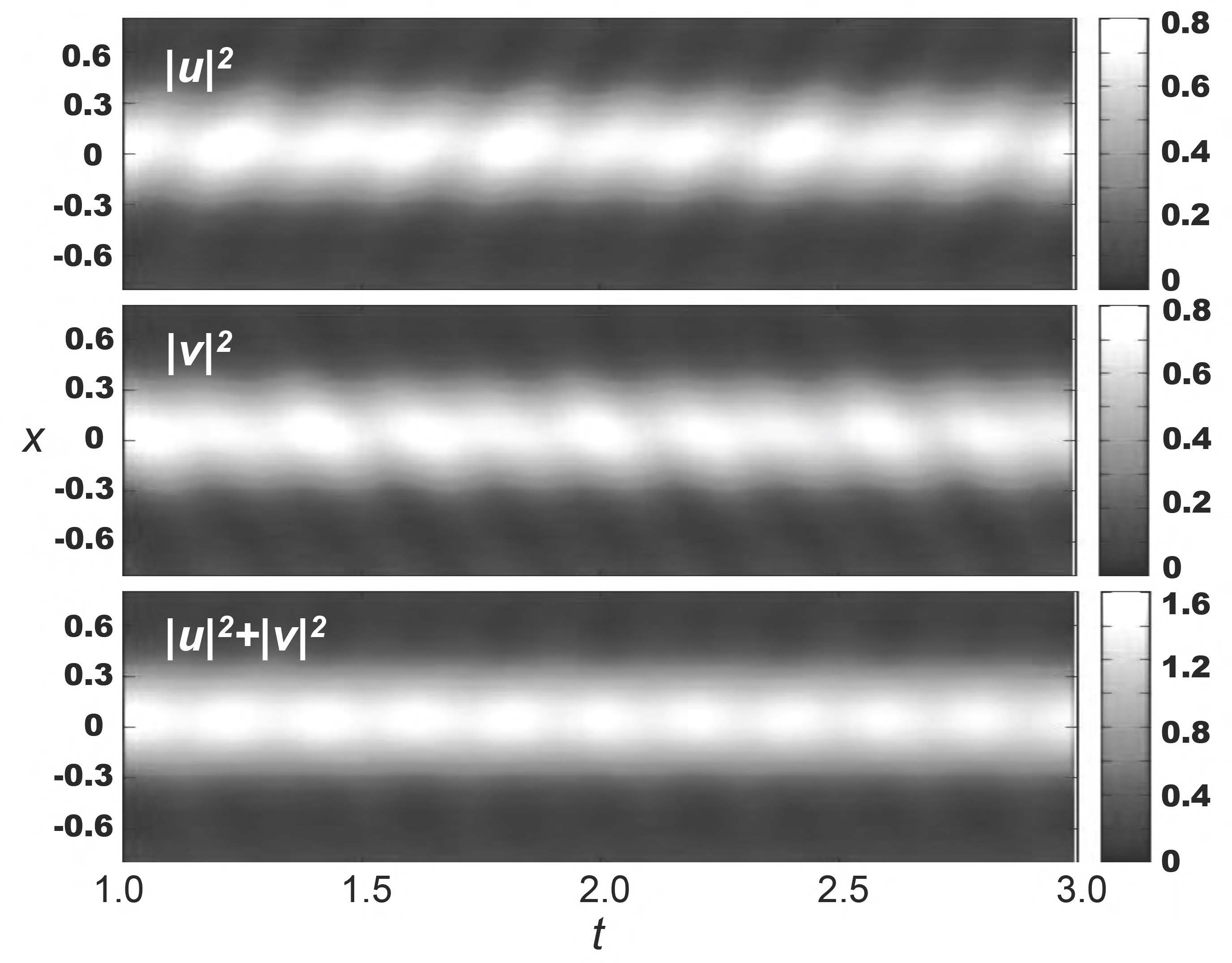}}
\caption{Density plots, $|u(x,t)|^2$, $|v(x,t)|^2$ and $|u(x,t)|^2+|v(x,t)|^2$, corresponding to the case 
shown in the lower set of three panels of Fig.~\ref{fig-15}, at positions $x$ for the time interval $1<t<3$. 
All quantities are in dimensionless units.
}
\label{fig-16}
\end{figure}

To conclude our study related to striped solitons and resonant interference effects, we present results obtained in 
longer time intervals, for the case that the unperturbed Raman is given by $\Omega_0=10$, with $k_L=4$, as in 
Fig.~\ref{fig-13}, but with a much smaller amplitude of the modulations, such that $\Omega_1=0.05\Omega_0$.  
The investigation of the interval of $\omega$ where interferences can be found is shown in Fig.~\ref{fig-14},
considering a small initial phase of oscillations between components given by $\phi=\pi/8$.
As shown by the set of five panels (for $1\le t\le 8$) with given values of $\omega$, resonant interference effects due to 
the perturbation are verified only in the interval $39>\omega >22$, with maxima interferences occurring for 
$\omega\sim 32$ (the middle panel, where we have also included with dashed line the unperturbed case, for comparison).
The results for $\omega>39$ and $<22$ are almost identical with the non-perturbed case, $\Omega_1=0$. 
Therefore, we select the case with $\omega=32$ to show in more detail the oscillating behavior, which is presented
in Figs.~\ref{fig-15} and \ref{fig-16}. In the lower panel of Fig.~\ref{fig-15}, we consider a larger time interval with  
$0\le t\le16$ (lower panel). The middle panel ($2\le t\le7$) serves to show the change in the frequency of the oscillations, 
such that for each two cycles another cycle is emerging, which can be verified for $10\le t\le16$.
In all the three given panels, for comparison we are including in dashed line the unperturbed case.

\section{High frequency modulations. Averaged GP equations}
In the case that we have rapidly and strongly varying Raman oscillations $\Omega(t)$, it is useful to derive 
the corresponding averaged GP equation, such that one can reduce the time-dependent modulated Raman
frequency to the constant one $\Omega_0$, by renormalizing the spin-orbit coupling and the non-linear parameters,
as we are going to show in this section. By matching the averaged results with the full-numerical ones, obtained with 
real-time evolution, we are also verifying numerically how fast and strong should be the time oscillations in order to 
validate the averaging approach.  
In order to derive the averaged over rapid modulations system of equations,
we first apply the following unitary transformation~\cite{SKMA,2013Zhang} in 
Eq.~(\ref{eq01}):
\begin{eqnarray}\label{eq16}
\Phi \equiv \left( \begin{array}{c} U \\ V \end{array} \right) 
&=&\left(
\begin{array}{cc}
\cos(q) & {\rm i}\sin(q) \\ {\rm i}\sin(q) & \cos(q)
\end{array}
\right)
\left( \begin{array}{c} u \\ v \end{array} \right) ,
\end{eqnarray}
where  $ q\equiv q(t) = (\Omega_1/\omega)\sin(\omega t)$ is given by the requirement that the time-dependent part of the
Raman frequency does not appear explicitly in the coupled equation for $\Phi$. 
When performing the time averaging of the interactions, together with the SOC parameter $k_L$, the parameters 
of the non-linear interaction have also to be renormalized. They are replaced by parameters that contains
zero-order Bessel function, considering that 
\begin{equation}\label{eq17}
\frac{1}{2\pi}\int_0^{2\pi} d(\omega t) \exp\left({\frac{{\rm i} n\Omega_1}{\omega}\sin(\omega t)}\right)
= J_0\left(\frac{n\Omega_1}{\omega}\right).
\end{equation}
Then, by defining $\chi\equiv 2\Omega_1/\omega$, the coupled equation, averaged over the period of rapid modulations, 
with $V_{tr}=0$, can be written as\cite{2013Zhang}:
{\small \begin{eqnarray}  \label{eq18}
&\mathrm{i} \displaystyle\frac{\partial\Phi}{\partial t}&
=\left[-\frac{1}{2}\frac{\partial^2 }{\partial x^2} -\mathrm{i}k_L J_0(\chi)
\sigma_z\frac{\partial }{\partial x} +\Omega_0\sigma_x \right]
\left( \begin{array}{c} U \\ V\end{array} \right) 
\\ 
&-&  \left( \begin{array}{cc} \alpha_+|U|^2 + \alpha_1|V|^2& \alpha_0U^*V \\ 
 \alpha_0V^*U&
 \alpha_-|V|^2 + \alpha_1|U|^2 \end{array} \right)
 \left( \begin{array}{c} U \\ V\end{array} \right),\nonumber
\end{eqnarray}}
where
\begin{eqnarray}\label{eq19}
\left. \begin{array}{l} 
\alpha_0 \equiv \left(\beta -\frac{1+\gamma}{2}\right)\frac{1-J_0(2\chi)}{4},\\ \\
\alpha_\pm\equiv \alpha_0 + \frac{1+\gamma}{2} \pm \frac{1-\gamma}{2} J_0(\chi),\\ \\
\alpha_1 \equiv \beta - 2\alpha_0 .  \end{array}\right\}
\end{eqnarray}

In the case of gauge symmetry, with $\beta = \gamma =1$, we  have $\alpha_0=0$ and $\alpha_{i\ne 0}=1$; i.e., the  
nonlinear part of the above coupling equation for ($U,V$) is exactly the same is the ones obtained for $(u,v)$, such that
the time averaging is only renormalizing the  SOC parameter $k_L$ to 
\begin{equation}
\kappa_{eff} =k_L J_0\left(\frac{2\Omega_1}{\omega}\right), \label{eq20}
\end{equation}
as one can verify by comparing the coupled Eqs.~(\ref{eq18}) with (\ref{eq02}).
This approach for tuning of the SOC parameter  has been  confirmed recently in an experiment reported in Ref.~\cite{2015Spielman}.
Therefore, when considering rapid variations of the Raman frequency, the spin-orbit coupling $k_L$ can be tuned in order to 
control the solitons in a BEC with SOC. In particular, it can be quite useful to transform striped solitons to regular solitons, and 
vice versa. With the appropriate ratio between amplitude $\Omega_1$ and frequency $\omega$ of the Raman oscillations, a
given value of $k_L$ for region II, where $ k_L^2 > \Omega_0$, can be changed to $ \kappa_{eff}^2 < \Omega_0$, where we obtain
regular soliton solutions, such that all the theory developed before (in sections II and III) for constant Raman frequency,  can be 
applied.  

The above can be exemplified by the results shown in Fig.~\ref{fig-03}, which are for regular soliton solutions, with $k_L$= 4 and  
$\Omega_0=$80 and  20, respectively. These results are for region I, but can also be applied to the case that we have originally
$k_L$ larger than $\Omega_0$, if the time modulation of the Raman frequency $\Omega(t)$, given by Eq.~(\ref{eq05}), 
is such that the ratio between $\Omega_1$ and $\omega$ will give us $\kappa_{eff}=k_LJ_0(\chi)=$4.
We could take initially, $k_L=10$, for example, as it is larger than $\Omega_0$ in both the cases shown in Fig.~\ref{fig-03} ,
with the parameters of the time-modulating Raman such that $J_0(\chi)=0.4$.

When considering other values for the nonlinear parameters, as a general remark we noticed that stable soliton solutions 
are obtained for attractive two-body interactions. Another remark, when considering the averaged approach, is that 
for $\beta \neq 1 $,  we can also have conditions with zero in the off-diagonal terms of the non-linear two-body matrix, 
which brings the Eq.~(\ref{eq18}) to the same format as Eq.~(\ref{eq02}). 
This happens for $\beta = (1+\gamma)/2$, with $\alpha_0=0$, $\alpha_1=\beta$, and $\alpha_\pm=\beta\pm(1-\beta)J_0(\chi)$.
 The more general cases, as  for $\alpha_0\ne 0$, or $2\beta \ne (1+\gamma)$,  new terms appear, corresponding to the effective 
 four-wave mixing ($\sim U^2V^{\ast}, V^2U^{\ast}$ ). These terms can lead to new possibilities, such as a way  
 to control the atom number oscillations between two components (internal Josephson effect\cite{Shenoy,Zhang2012}).

\subsubsection{The solitonic solutions}
The solitonic solutions for the averaged GP equations can be found by applying the multi-scale method~\cite{AFKP} to the  
two regions defined by the linear spectrum, which are given by Eq(\ref{eq03}).
By using this multi-scale method, for values of the chemical potential near the  bottom of the dispersive curve, 
with $\mu = -\Omega_0 -\varepsilon^2 w_0$ ($\varepsilon \ll 1$),  $\omega_0$ is the free parameter,  in region I (see Fig.~\ref{fig-01}), we obtain   
\begin{eqnarray}
u_s^{(I)} &=& \varepsilon\sqrt{\frac{2w_0}{\alpha_++\alpha_1+\alpha_0}}\mbox{sech}\left( \varepsilon\sqrt{\frac{2w_0}{\Delta_{eff}}}x \right)
,  \label{eq21} \\
v_s^{(I)}&=&-u_s^{(I)},\;\;\; \Delta_{eff}=1-\frac{\kappa_{eff}^2}{\Omega_0}.\label{eq22}
\end{eqnarray}
In the region II,  where $\kappa_{eff}^2 > \Omega_0$ and two minima exist in the momentum space, we can take the chemical potential 
as $\mu=w_{min}-\varepsilon^2w_0$ (see Fig.~\ref{fig-01}), with 
\begin{equation}\label{eq23}
w_{min} =\frac{1}{2}\kappa_{0}^2 - \kappa_{eff}^2, \;\; \kappa_0\equiv\pm\sqrt{\kappa_{eff}^2-\Omega_0^2/\kappa_{eff}^2}
\end{equation}
and look for solutions of the form $(u,v)=\varepsilon (A, B) \exp(\pm i  \kappa_0 x)$.
For the result, we obtain a bright soliton solution with the form given by
{\small \begin{equation}\label{eq24}
\left( \begin{array}{c} 
u_{s}^{(II)}\\ 
v_{s}^{(II)}\end{array} \right)
= \left( \begin{array}{c} 
\Omega_0\\ 
-\kappa_{eff}(\kappa_{eff}\pm \kappa_0)
\end{array} \right)
\frac{\varepsilon f(x)e^{\pm i  \kappa_0 x-i\mu t)}} {\sqrt{|\kappa_{eff}\pm \kappa_0|}}
\end{equation}}
where
\begin{eqnarray}\label{eq25}
f(x)&=&\frac{\sqrt{2w_0 \kappa_{eff}}}{\sqrt{\alpha_+(\kappa_{eff}^4+
\kappa_{eff}^2\kappa_0^2)+(\alpha_1+\alpha_0)\Omega_0 ^2}}\times,\nonumber\\
&\times&\mbox{sech}\left(\varepsilon\sqrt{\frac{2w_0 \kappa_{eff}^2}{\kappa_0^2}} x \right)
.\end{eqnarray}

Analogically, the striped soliton solution can be found as linear superpositions of solutions represented by
Eq.(\ref{eq24}).
As already known, these solutions are used to describe the  longitudinal and transversal spin polarizations of the
solitons~\cite{AFKP}, with
\begin{eqnarray}
\langle\sigma_z\rangle&=&\frac{1}{N}\int_{-\infty}^{\infty}dx(|u|^2-|v|^2),
\nonumber\\
 \langle\sigma_x\rangle&=&\frac{1}{N}\int_{-\infty}^{\infty}dx(u^{\ast}v + uv^{\ast}),\label{eq26} \\
 N&\equiv&\int_{-\infty}^{\infty}dx(|u|^2+|v|^2).\nonumber
 \end{eqnarray}
In the region I, where $\Omega_0 <  \kappa_{eff}^2$, 
the solitons are fully polarized along the $x-$axis. The same approach is valid for stripe solitons in the region II.
However, the polarization along $z$ is not zero for solitons with momentum $k=\pm \kappa_{0}$. From 
Eqs.~(\ref{eq21})-(\ref{eq25}), we obtain 
\begin{equation}\label{eq27}
\langle\sigma_z\rangle^{(II)}= -\sqrt{1-\Omega_0^2/\kappa_{eff}^4},\;\;
\langle\sigma_x\rangle^{(II)}=-\frac{\Omega_0}{\kappa_{eff}^2}.
\end{equation}
Thus, by varying the ratio $\chi$, and so $\kappa_{eff}$, we can observe quantum phase transition in the pseudo-spin polarization 
$\langle\sigma_z\rangle^{(II)}$ of the soliton.  These results for the soliton polarization are analogous to the ones obtained for 
the repulsive BEC  in the framework of the Dicke model in \cite{2013Zhang}.

\begin{figure}[h]
\includegraphics[width=8.5cm]{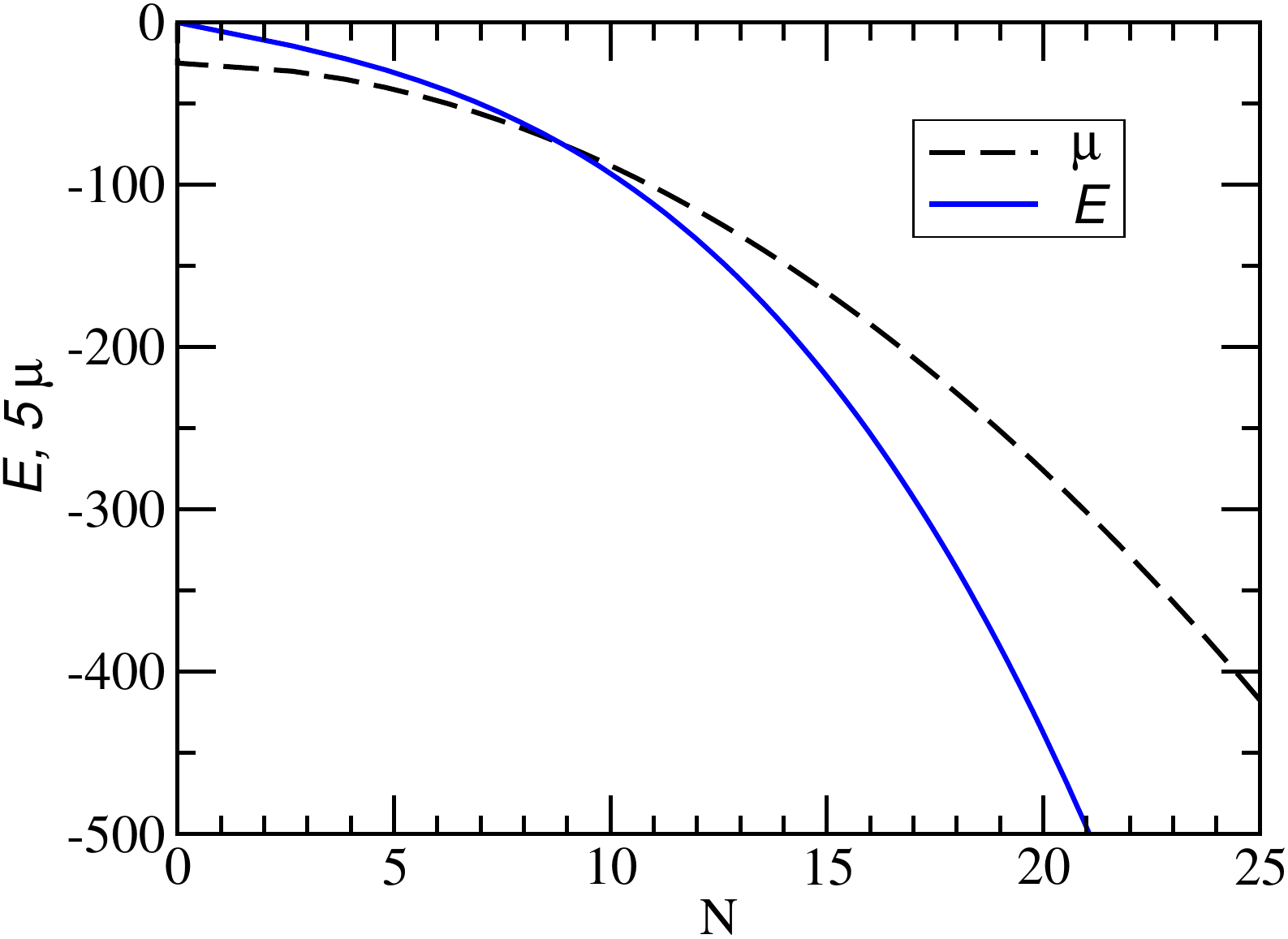}
\caption{ 
The energy $E$ (solid line) and chemical potential $\mu$ (dashed line) are shown as functions of the atom
number $N$, by considering the parameter $\beta=\gamma=1$, $k_L=$8,  $\Omega_0=0$ and 
$\chi=2\Omega_1/\omega=$3.7152 [$J_0(\chi)= -0.4$, $\kappa_{eff}=-3.2$], with $\varepsilon^2w_0=0.4$.
All quantities are in dimensionless units.
}
\label{fig-17}
\end{figure}

\begin{figure}[h]
\includegraphics[width=8.5cm]{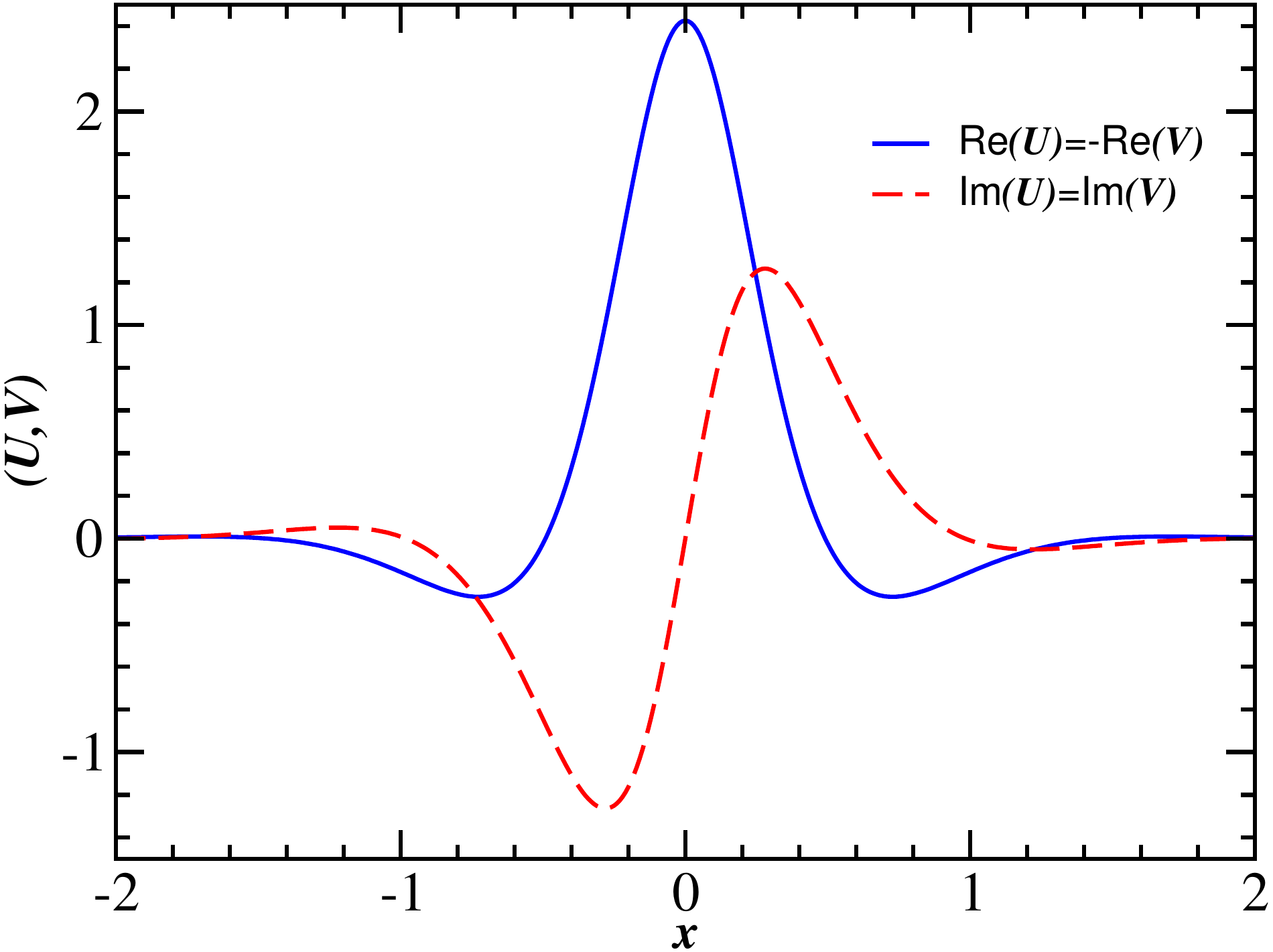}
\caption{
The real (solid-blue line) and imaginary (dashed-red) parts of the wave-function components are shown
for $E=-48.6$, $N=6.86$, $\mu=-11$, with the other parameters as in Fig.~\ref{fig-17}.
All quantities in dimensionless units.
} \label{fig-18}
\end{figure}

With the understanding that the results obtained in this section are valid in a more general context for constant 
values of the Raman frequency,  with Fig.~\ref{fig-17} we are showing the dependence of the energy and chemical 
potential on the number of atoms $N$ for the case that $\Omega_0=0$, $\beta=\gamma=1$, $k_L=8$, when considering 
$\chi=2\Omega_1/\omega=3.7152$ (with both $\Omega_1$ and $\omega$ very large), which give us $J_0(\chi)=-0.4$. 
Note that, in this simple case we have for the dispersion relation (\ref{eq03}) $w_\pm(k) = k(k/2\pm k_L)$, with the signal 
of the give SOC moving from the original $k_L=8$ to a negative one, $\kappa_{eff}=-3.2$. Threfore,  after considering the 
time averaging, in this particular case with $\Omega_0=0$, we obtain $w_\pm(k) = k(k/2\pm \kappa_{eff})$, such that 
both $w_+$ and $w_-$ have the same shape as $w_-$ shown in Fig.~\ref{fig-01}, but with minima given at 
$k=-\kappa_{0}$ (for $w_+)$; and $k=\kappa_{0}$ (for $w_-$).
As we are in region II, even after the averaging, the soliton solutions are not regular ones, being expected to have shapes
with some oscillations. In Fig.~\ref{fig-18}, we are illustrating the kind of solutions we obtain, by presenting the real 
and imaginary parts of the wave-function components when considering the particular case with $N=6.86$, $E=-48.6$ and 
$\mu=-11$. 

\subsubsection{Full numerical versus averaged results}
To conclude this section, we are comparing the time evolution results obtained with the effective
time-averaging approach (where the SOC parameter is $\kappa_{eff}$), with the ones obtained in real time,
with the SOC parameter $k_L$ and explicit Raman frequency modulated by $\Omega_1\cos(\omega t)$. 
\begin{figure}[tbph]
\centerline{
\includegraphics[width=8cm]{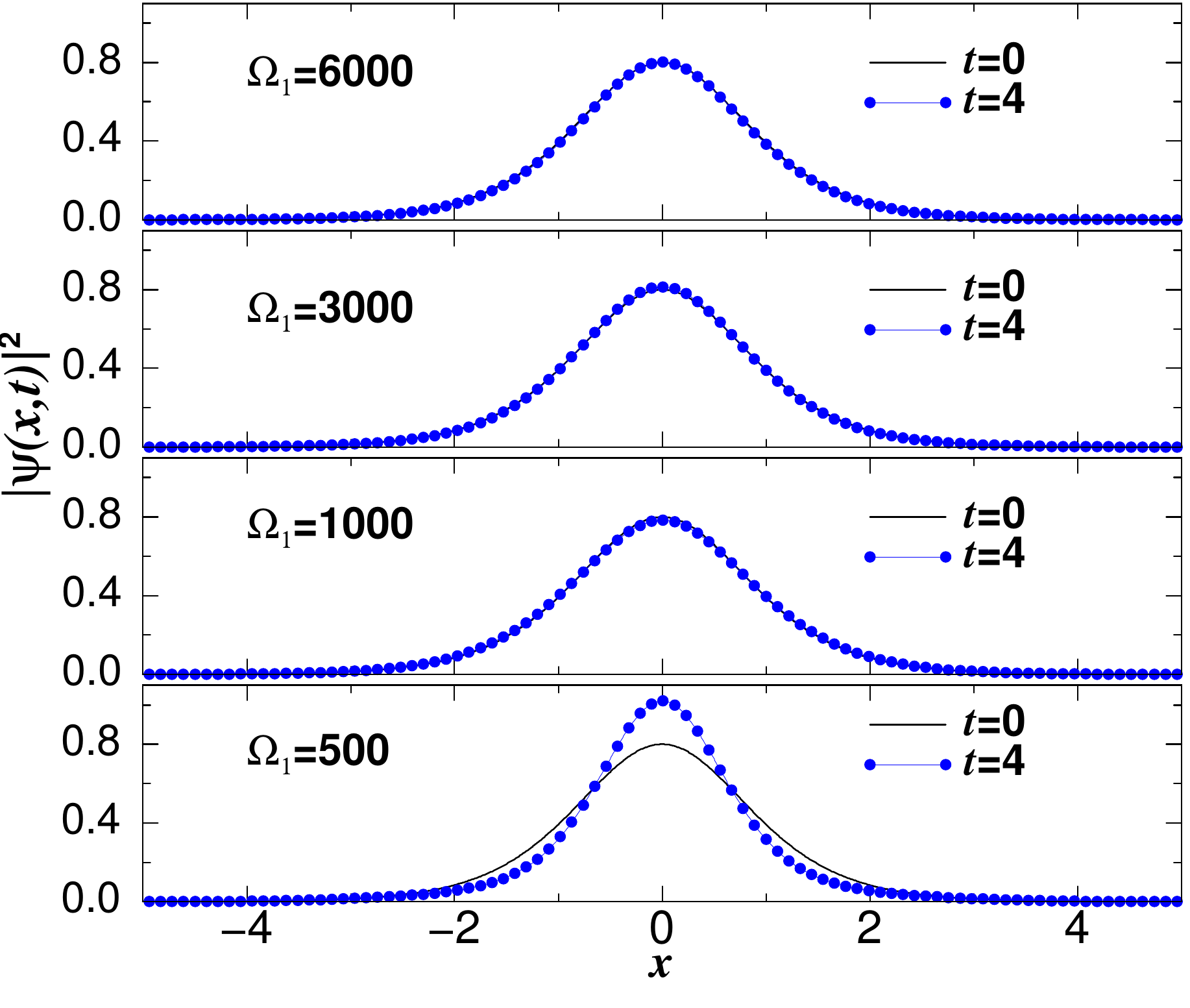}}
\caption{Real-time evolution of the soliton profile, for $\Omega_0=120$, $k_L=8$, with 
$\beta=\gamma=1$,  considering several values of $\Omega_1$ (shown inside the panels) and $\omega$,
such that $J_0(\chi)=0$ ($\chi=2.4$).
At $t=0$, the results in each of the panels coincide with the ones obtained with the averaged formalism 
(where $\kappa_{eff}=0$). At larger times (represented by $t=4$), we show that the 
agreement between real-time and averaged results is improved as one increases $\Omega_1$. 
The other parameters are $\beta=\gamma=1$. All quantities are in dimensionless units.}
\label{fig-19}
\end{figure}

\begin{figure}[h]
\centerline{
\includegraphics[width=8cm]{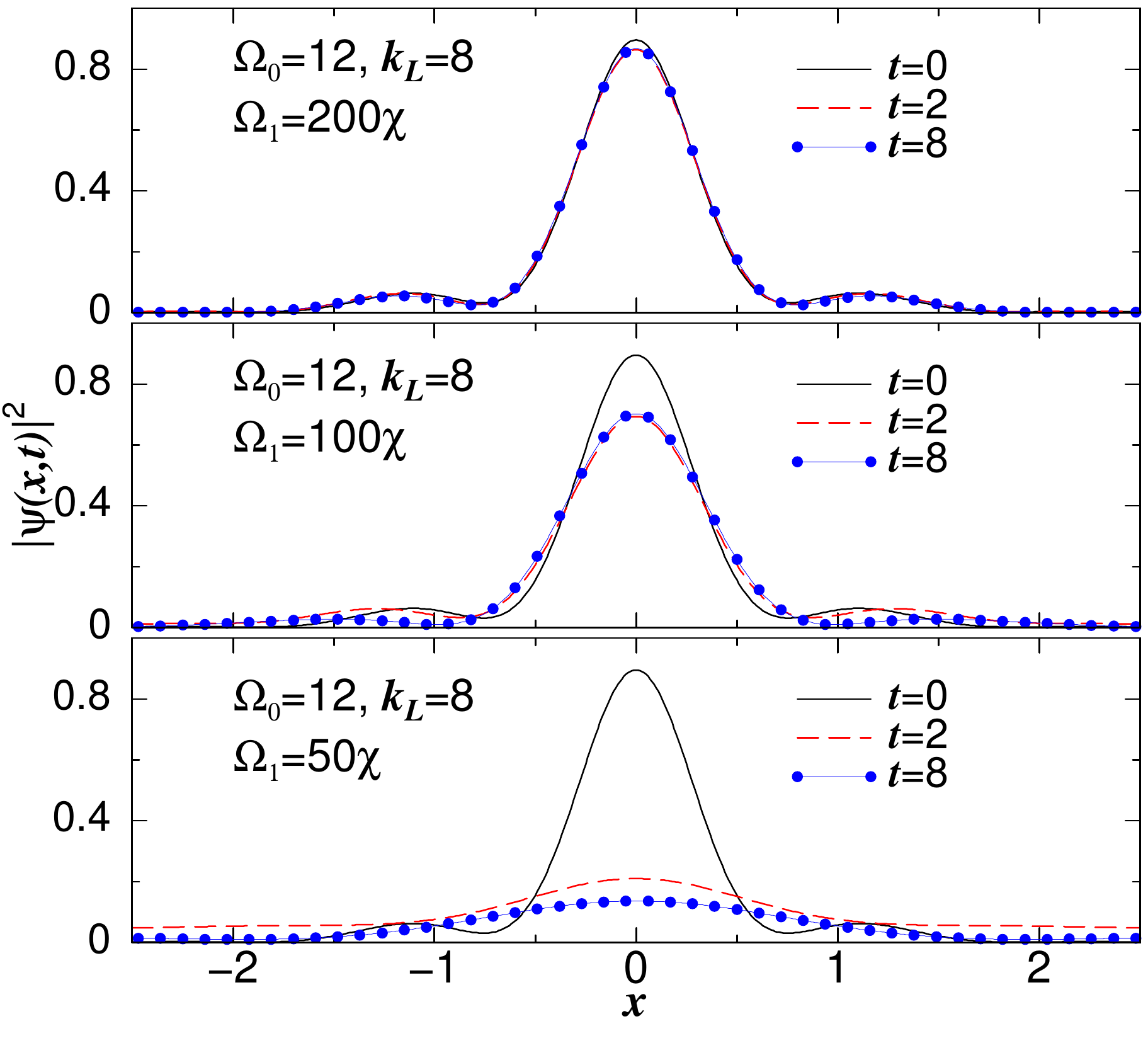}}
\caption{(color on-line) Evolution of the soliton profile, for $\Omega_0=12$, $k_L=8$, with $\beta=\gamma=1$,
in three panels for different values of $\Omega_1$, with  $\chi=2\Omega_1/\omega=1.52$ (such that $J_0(\chi)=1/2$),
implying that $\kappa_{eff}=4$. In this case, we have $\kappa_0=2.65$. 
All quantities in dimensionless units.
}
\label{fig-20}
\end{figure}

\begin{figure}[h]
\centerline{
\includegraphics[width=8.5cm]{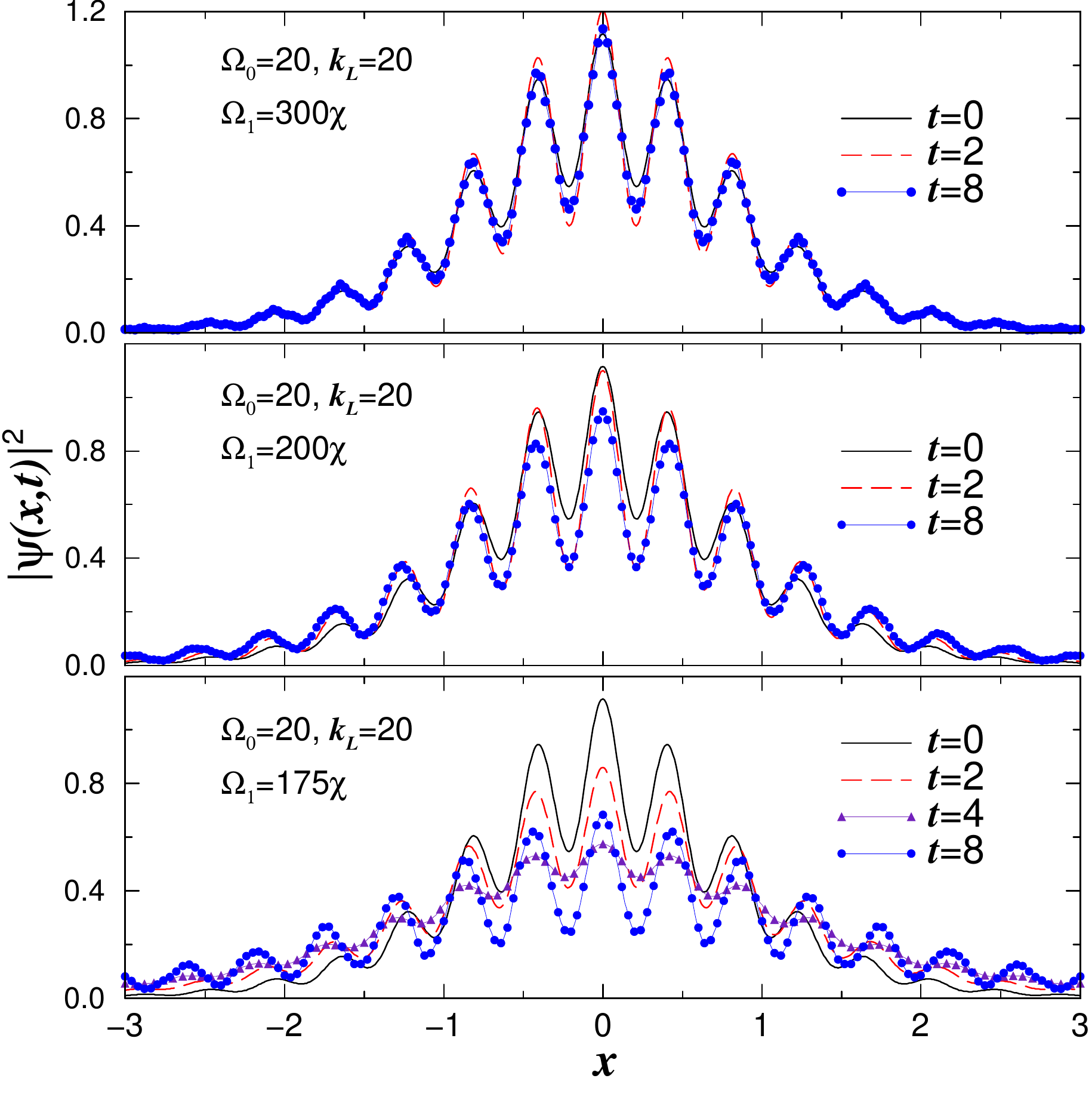}}
\caption{(color on-line) The evolution of striped soliton profiles are shown in three panels, for $\Omega_0=20$ and $k_L=20$, 
considering a few values of $\Omega_1$ and $\omega$ in the real time propagation. 
For $t=0$, the results are the same ones obtained by the averaged formalism, where $\kappa_{eff}=-8$. 
In all the frames, the ratio between the parameters $\Omega_1$ and $\omega$ is fixed, given by 
$\chi=2\Omega_1/\omega=3.7152$, implying in $J_0(\chi)=-0.4$. In this case, we have $\kappa_0=7.60$.
(All quantities in dimensionless units)}
\label{fig-21}
\end{figure}

With Figs.~\ref{fig-19}, \ref{fig-20} and \ref{fig-21}, we are exemplifying our results for the comparison of 
averaged results with real-time dependent numerical simulations. All the results for the time-averaged formalism
that are shown in these examples are verified to be numerically very stable in the time evolution.

The results given in Fig.~\ref{fig-19} are for the region I, with $\Omega_0=$120 ($>k_L^2$), by considering the SOC 
parameter $k_L=8$. For the
time-dependent Raman frequency we assume $\Omega_1$ and $\omega$ such that $\chi\sim 2.4$, 
implying that $J_0(\chi)=0$. Therefore, in this case, the averaged SOC parameter is $\kappa_{eff}=0$. 
As shown in the four panels, the averaged results present good agreement with the real-time simulations 
when $\Omega_1$ is about 10 times or more larger than $\Omega_0$. 

For the region II, we are illustrating with Figs.~\ref{fig-20} and \ref{fig-21}, for two quite different combinations
of Raman frequency and SOC parameters.

In Fig.~\ref{fig-20}, we present our results obtained for $k_L=8$ and $\Omega_0=$12, with
$\Omega_1$ and $\omega$ such that $\chi\sim 1.52$.  
As $J_0(\chi)=1/2$ and $\kappa_{eff}=4$, we are in region II ($\Omega_0<\kappa_{eff}^2$).
The results are shown in three panels, for different values of $\Omega_1=50\chi=76$ (lower panel), 
$=100\chi=152$ (middle panel) and $=200\chi=304$ (upper panel). 
The parameters used in this case correspond to one of the examples presented in Fig.~\ref{fig-09} for 
Josephson oscillations, where we have constant $\Omega_0=12$, with $k_L=4$. We should note that
the striped solitons shown in Fig.~\ref{fig-20} have the main maximum at the center, with only 
one pair of maxima visible in each side, due to the choice of parameters which are close to the border 
between the regions for striped and regular solitons.

In Fig.~\ref{fig-21}, we are considering a case where the effective SOC becomes negative, and we are more
deeply inside the region II. By departing from a large value of $k_L=20$, with the combinations of $\Omega_1$ 
and $\omega$, such that by fixing $\chi=3.7152$ and $J_0(\chi)=-0.4$, we have $\kappa_{eff}=-8$. The results are shown 
in three panels. For comparison, in all the three panels we also present the averaged results, which is equal to the unperturbed 
case with $t=0$. From our study of this case, we should also observe that for smaller values of $\Omega_1$ the 
real-time solutions become unstable, collapsing in a short time interval. The real-time solutions shown in the 
lower panel, for $\Omega_1=175\chi$, are already indicating this instability. When considering 
 $\Omega_1=150\chi$, the solution was already collapsed even at $t=2$.

By considering our results, exemplified by Figs.~\ref{fig-19}, \ref{fig-20} and \ref{fig-21}, as a general remark for 
the case of fast-time oscillations in the Raman frequency, our conclusion is that good agreements between the 
averaged results with the full-numerical simulations can be verified only for $\Omega_1$ about 10 times larger 
than $\Omega_0$ (where the frequency $\omega$ is of the order of $\Omega_1$), which is an approximate 
minimal condition for the time-modulations in the Raman frequency in order to keep stable the soliton solutions
during time evolution.  

In the next section, we summarize this work with our main conclusions.
 
\section{Conclusions}
In the present work, we have studied the existence and dynamics of solitons in Bose-Einstein condensates (BEC) with spin-orbit 
coupling (SOC) and attractive two-body interactions, by considering two coupled atomic pseudo-spin components with general
time-dependent Raman frequency, which can be constant, slowly or rapidly modulated in time.  For that, after defining the two
possible regions where two different kind of soliton solutions exist, regular or striped bright solitons, we first consider the Raman
frequency varying slowly and linearly in time, such that we can study the transition between the two kinds of soliton solutions;
from regular to striped ones and vice-versa. The regions are established by the relation between the SOC parameter $k_L$
and the constant part of the Raman frequency, $\Omega_0$, such that we have regular solitons in region I, when
$\Omega_0>k_L^2$; and striped solitons in region II, for $\Omega_0<k_L^2$. 
Next,  we study the internal Josephson oscillations between the atom numbers in 
soliton components, which are controlled by constant or periodically time-oscillating Raman parameter.
Different parameter configurations are studied for SOC in BEC, with parametric resonances  indicating a mechanism to 
control the soliton parameters, as well as the evolution of the solitons center of mass.  As shown, we are also presenting 
a variational analysis, valid particularly in the case that we obtain regular bright soliton solutions. The full-numerical simulations 
have confirmed the corresponding predictions.

In the limit of high frequencies, the system is described by a time-averaged Gross-Pitaevskii formalism 
with renormalized nonlinear and SOC parameters and additional modified phase-dependent nonlinearities. 
Therefore, by comparing full-numerical simulations with averaged results, we have studied the lower limits for the 
frequency of the Raman oscillations, in order to obtain stable soliton solutions. The results are shown in a few examples,
for both regions I and II.
One should note that, due to the normalization of the nonlinear interactions, new terms can emerge in the nonlinear coupling
of the averaged system for BEC with tunable SOC, when comparing with the original non-averaged formalism.
Corresponding to {\it the phase depending nonlinear coupling}, we have a new term $\sim\alpha_0$ appearing in the
off-diagonal matrix terms of the nonlinear coupling. This term can play important role for non-stationary processes in BEC with 
SOC, as well as in the Josephson oscillations between components of solitons with nonzero phase differences. 
This matter requires further separate investigation.

The expected relevance of the present study can be by predicting some effects, as well as in
the corresponding parameter control, in a possible  BEC experiment, such as in $^{7}$Li with attractive interatomic interactions,  
where the SOC can be engineered as an 
effective two-level atoms by an uniform magnetic field $B$ with two Raman laser beams. In this example, we have the 
linear transverse trap frequency, $\omega_{\perp}/(2\pi)=$1 kHz, the number of atoms $N=10^3$ and the wavelength of 
the  Raman lasers given by $\lambda=804$nm. Therefore, the Raman frequency can vary in the interval 
$(0.1-10) E_L$, where $E_L =\hbar^2 k_L^2/2m$ is the recoil energy and $k_L=2\pi/\lambda$. For $\Omega_0=0.1 E_L/\hbar$ 
we obtain $\Omega_0=2\pi \times 30$ kHz. Then the frequency of modulations are: for the resonant case the dimensionless 
$\omega=60$ corresponds to  $\omega = 2\Omega_0=60$ kHz, for the high frequency limit $\omega =300$ to  
$\omega = 10\Omega_0=300$ kHz.

\acknowledgements
The authors acknowledge partial support from Conselho Nacional de Desenvolvimento Cient\'\i fico e Tecnol\'ogico
[research-grants (LT and AG) and Proj. 400716/2014-3 (FKA, MB, LT)], 
Funda\c c\~ao de Amparo \`a Pesquisa do Estado de S\~ao Paulo [Projs. 2017/05660-0 (LT), 2016/17612-7 (AG)], and
Coordena\c c\~ao de Aperfei\c coamento de Pessoal de N\'\i vel Superior [program PVS-CAPES/ITA (LT)]. 
FKA is also grateful to IFT-UNESP for providing local facilities.

\end{document}